\documentclass[10pt,sigconf,letterpaper]{acmart}

\usepackage{booktabs} %
\usepackage{graphicx}   %
\usepackage{mathtools}
\usepackage{amsmath}
\usepackage{color, colortbl}
\usepackage[nolist]{acronym}
\usepackage{multirow}
\usepackage{euscript}
\usepackage{adjustbox}
\usepackage{array}
\usepackage{appendix}
\usepackage[tableposition=bottom]{caption}
\usepackage{subfig}
\usepackage{placeins}
\usepackage[noabbrev]{cleveref}

\usepackage{hyperref}
\def\Snospace~{\S{}}

\hypersetup{
    colorlinks,%
    citecolor=[rgb]{0.7,0,0},
    filecolor=[rgb]{0.2,0.2,0.2},
    linkcolor=[rgb]{0,0,0.7},
    urlcolor=[rgb]{0.7,0,0},
}

\definecolor{Gray}{gray}{0.9}

\newcolumntype{C}{>{\centering\arraybackslash}m{4.5em}}
\newcolumntype{D}{>{\centering\arraybackslash}m{3.5em}}
\newcolumntype{M}[1]{>{\centering\arraybackslash}m{#1}}
\newcolumntype{P}[1]{>{\centering\arraybackslash}p{#1}}
\newcolumntype{N}{@{}m{0pt}@{}}

\setcopyright{none}

\renewcommand\footnotetextcopyrightpermission[1]{} %

  \acmDOI{}
  \acmISBN{}
  \acmConference{Technical Report}{Mar 2023}{Marina del Rey, CA, USA}
  \acmYear{}
  \copyrightyear{2023}
  \acmArticle{}
  \acmPrice{}

\makeatletter
  \newcommand\EatSpacesHack{\@bsphack\@esphack}
\makeatother

  \renewcommand\comment[1]{\EatSpacesHack}
  \newcommand\yp[1]{\EatSpacesHack}
  \newcommand\gb[2]{\EatSpacesHack}
  \newcommand\reviewfix[1]{\EatSpacesHack}

  \newcommand\PostSubmission[1]{\EatSpacesHack}

  \makeatletter
    \renewcommand\section{\@startsection{section}{1}{\z@}%
      {-.5\baselineskip \@plus -2\p@ \@minus -.2\p@}%
      {.2\baselineskip}%
      {\ACM@NRadjust\@secfont}}
    \renewcommand\subsection{\@startsection{subsection}{2}{\z@}%
      {-.4\baselineskip \@plus -2\p@ \@minus -.2\p@}%
      {.18\baselineskip}%
      {\ACM@NRadjust\@subsecfont}}
    \renewcommand\subsubsection{\@startsection{subsubsection}{3}{\z@}%
      {-.35\baselineskip \@plus -2\p@ \@minus -.2\p@}%
      {-2\p@}%
      {\ACM@NRadjust{\@subsubsecfont\@adddotafter}}}
   \makeatother
  \newcommand{\RelaxFloats}{
  	\renewcommand{\topfraction}{0.9}
  	\renewcommand{\floatpagefraction}{0.9}
  	\renewcommand{\textfraction}{0.1}
  }
  \RelaxFloats

\acrodefplural{AS}[ASes]{Autonomous Systems}
\acrodefplural{VP}[VPs]{Vantage Points}
\acrodef{NAT}[NAT]{Network Address Translation}
\acrodef{CG-NAT}[CG-NAT]{Carrier-Grade NAT}
\acrodefplural{CDN}[CDNs]{Content Delivery Networks}
\acrodef{RIR}[RIR]{Regional Internet Registry}
\acrodefplural{RIR}[RIRs]{Regional Internet Registries}

\begin{document}
\title{Reasoning About Internet Connectivity}

  \setlength{\dblfloatsep}{-5pt}
  \addtolength{\abovecaptionskip}{-4pt}
  \setlength{\dbltextfloatsep}{0pt}
  \setlength{\abovedisplayskip}{1.5pt plus 1pt minus 1pt}
  \setlength{\belowdisplayskip}{1.5pt plus 1pt minus 1pt}

  \settopmatter{authorsperrow=4}
  \author{Guillermo Baltra}
  \affiliation{%
    \department{USC/ISI}
    \city{Marina del Rey}
    \state{California}
    \country{USA}
  }
  \email{baltra@ant.isi.edu}

  \author{Tarang Saluja}
  \affiliation{%
    \department{Swarthmore College}
    \city{Swarthmore}
    \state{Pennsylvania}
    \country{USA}
  }
  \email{tsaluja1@swarthmore.edu}

  \author{Yuri Pradkin}
  \affiliation{%
    \department{USC/ISI}
    \city{Marina del Rey}
    \state{California}
    \country{USA}
  }
  \email{yuri@isi.edu}

  \author{John Heidemann}
  \affiliation{%
    \department{USC/ISI and Thomas Lord Dept.~of CS}
    \city{Marina del Rey}
    \state{California}
    \country{USA}
  }
  \email{johnh@isi.edu}

  \renewcommand{\shortauthors}{Baltra et al.}

\begin{abstract}
Innovation in the Internet
  requires a global \emph{Internet core}
  to enable communication between
  users in ISPs and services in the cloud.
Today, this Internet core is
  challenged by \emph{partial reachability}:
  political pressure threatens fragmentation by nationality,
  architectural changes such as carrier-grade NAT make connectivity conditional,
  and operational problems and commercial disputes make
  reachability incomplete for months.
We assert that \emph{partial reachability is a fundamental part of the Internet core}.
While other studies address partial reachability,
  this paper is the first to
  \emph{provide a conceptual definition of the Internet core}
  so we can reason about  reachability from principles first.
Following the Internet design,
  our definition is guided by reachability, not authority.
Its corollaries
  are \emph{peninsulas}:  persistent regions of partial connectivity;
  and \emph{islands}: when networks are partitioned
    from the Internet core.
We show that the concept of \emph{peninsulas and islands can improve
  existing measurement systems}.
In one example, they show that
  RIPE's DNSmon suffers misconfiguration
  and persistent network problems
  that are important,
  but risk obscuring
  operationally important connectivity changes
  because they are $5\times$ to $9.7\times$ larger.
Our evaluation also informs policy questions,
  showing
  no single country or organization can unilaterally control the Internet core.
\end{abstract}

\begin{acronym}[AS]
\acro{AS}{Autonomous System}
\acro{RDNS}{Reverse DNS}
\acro{VP}{Vantage Point}
\end{acronym}

\maketitle

\vspace*{-1ex}

\section{Introduction}
	\label{sec:introduction}

Today's Internet is very different from when it was created.
In 1980, Postel defined ``an internet'' as
  ``a collection of interconnected networks'',
  such as the ARPAnet and X.25~\cite{Postel80b}.
In 1995,
  the Federal Networking Council defined ``Internet''
  as (i) a global address space,
  (ii) supporting TCP/IP and its follow-ons,
  that (iii) provides services~\cite{nitrd}.
Later work added DNS~\cite{IAB20a} and IPv6.
Yet today,
  users at home and work access the Internet indirectly through
  \ac{NAT}~\cite{Tsuchiya93a},
  or from mobile devices
  through \ac{CG-NAT}~\cite{Richter16c}.
Many public services operate from the cloud,
  visible through rented or imported IP addresses,
  inside networks with multiple levels of virtualization~\cite{Greenberg09a}.
Media is replicated in \acp{CDN}.
Access is mediated by firewalls.
Yet to most users, Internet services are so seamless
  that technology recedes
  and the web, Facebook, or their phone is their ``Internet''.

\textbf{The Core and Its Challenges:}
Today's rich, global Internet services exist
  because \emph{the Internet architecture
  created
  a single, global \textbf{Internet core} where all can communicate freely}.
The Internet core has enabled %
  40 years of  permissionless
  innovation~\cite{Lemley01a},
  from e-mail and remote login to the web and streaming media,
  today's Internet powers telephony and video distribution.
Although today many edge devices are client-only,
  \emph{continued innovation and international exchange depends on a
  near-completely connected Internet core}.

But today \emph{universal reachability in Internet core is often challenged
  with threats of fragmentation}. %
\emph{Political} pressure pushes to Balkanize the Internet
  along national borders~\cite{Drake16a,Drake22a}.
Consdier
  Russia's 2019 sovereign-Internet law~\cite{BBC19a,RBC21a,Reuters21a}
and
  national ``Internet kill switches''
  debated in U.S.~\cite{GovTrack20a}, the U.K.,
  and deployed elsewhere~\cite{Cowie11a,Coca18a,Griffiths19a,Taye19a}. %
We suggest that \emph{technical clarification can inform policy discussions}
  as threats of de-peering
  place the global Internet at risk (\autoref{sec:policy_applications}).

\emph{Architectural} pressures from 30 years of evolution
  segment today's Internet core:
  services are gatewayed through proprietary cloud APIs,
  users are usually second-class and client-only due to \ac{NAT},
  firewalls interrupt connectivity,
  and the world straddles a mix of IPv4 and IPv6.
Architecture sometimes follows politics,
  with
  China's Great Firewall managing their
  international communication~\cite{Anonymous12a,Anonymous14a},
  and
  Huawei proposing ``new Internet'' protocols~\cite{huawei2020}.
We suggest that technical methods
  can help us \emph{reason about changes to Internet architecture},
  both to
  understand the implications of partial address reachability
  and evaluate the maturity of IPv6.

\emph{Operational} challenges
  can cause partial reachability.
Peering disputes can cause long-term partial reachability~\cite{ipv6peeringdisputes}.
Unreachability has been recognized and detected experimentally~\cite{Dhamdhere18a},
  and systems exist that mitigate partial
  reachability~\cite{andersen2001resilient,katz2008studying,katz2012lifeguard}.
Yet an understanding how duration and breadth of partial reachability
  has remained elusive.

\textbf{Contributions:}
Our first contribution that
  \emph{an Internet core},
  the global address space
  to which everyone interconnects,
  is essential to continued innovation.
\emph{Understanding partial connectivity
  is key to reasoning about challenges} to the Internet and innovation.
We hope that recognition
  the importance of one, global, open Internet core
  will help clarify the stakes
  when nations assert sovereignty
  and architectural changes require mediated communication.
We also show that \emph{peninsulas},
  regions of partial connectivity that are sometimes long-lasting,
  are an real-world problem as serious as network outages~\cite{Schulman11a,quan2013trinocular,Shah17a,richter2018advancing,guillot2019internet}.

Our second contribution is to offer
  a rigorous, conceptual definition of the Internet core as
  \emph{the strongly connected component of more than 50\% of active,
  public-IP addresses that can initiate communication with each other}
  (\autoref{sec:definition}).
By requiring bidirectional initiation,
  this definition
  captures the uniform, \emph{peer-to-peer nature
  of the Internet core}
  necessary for first-class services.
The 50\% requirement defines \emph{one, unique} Internet core,
  without central authority, historical precedent, or special locations,
  since multiple majorities
  are impossible.
Unlike prior work~\cite{andersen2001resilient,katz2008studying,katz2012lifeguard}, this \emph{conceptual} definition
  avoids dependence on any specific measurement system.
We have realized this definition in operational systems
  with two different data sources (\autoref{sec:dnsmon} and
  elsewhere~\cite{Baltra23a}).
This conceptual Internet core
  defines an asymptote against which our current and future operational systems can compare.

Our final contribution is to \emph{use our definition
  to clarify policy, architectural, and operational questions}.
We bring technical light to policy choices
  around national networks (\autoref{sec:policy_applications}) and de-peering (\autoref{sec:internet_partition}).
Our definition can help evaluate the IPv4/v6 transition
  and clarify operational questions in IPv4 address use,
  in outage detection~\cite{Schulman11a,quan2013trinocular,Shah17a,richter2018advancing,guillot2019internet},
  and WAN~\cite{andersen2001resilient,katz2008studying,katz2012lifeguard} and cloud~\cite{Schlinker17a} reachability optimization
  (\autoref{sec:arch_applications}).
We apply our results to widely used RIPE DNSmon (\autoref{sec:dnsmon}).
Today DNSmon shows
   persistent high query loss (5--8\% to the DNS Root~\cite{RootServers16a}),
   we show that most of this loss is due to misconfiguration
   and
   persistent partial connectivity.
While such factors matter,
  they are $5\times$ and $9.7\times$ (IPv4 and v6) larger than
  other operationally important signals.
Separating them therefore improves sensitivity in DNSmon~\cite{Amin15a} (\autoref{sec:dnsmon}).

\textbf{Artifacts and ethics:}
All of the data used %
  and
    created~\cite{ANT22b}
  in this paper
  is available at no cost.
Our work poses no ethical concerns:
  we reanalyze existing data with new algorithms,
  and have no information about individuals.
IRB review declared our work non-human subjects
  research
  (USC IRB IIR00001648).

\section{Problem: Partial Reachability}
	\label{sec:problem}

To understand Internet connectivity
  we must rigorously define \emph{the Internet core}
  to which we connect,
  to answer the political, architectural, and operational questions
  from \autoref{sec:introduction}.

First, a definition should be both
  \emph{conceptual} and \emph{operational}~\cite{scientific_methods}.
Our conceptual definition in \autoref{sec:definition}
  articulates what we would \emph{like} to observe
  and suggests a limit that an implementation can approach.
In \autoref{sec:dnsmon} we operationalize our definitions
  to improve understanding of DNSmon.
Prior definitions are too vague to operationalize.

Second, a definition must give both sufficient \emph{and}
  necessary conditions to be part of the Internet core.
Prior work gave only sufficient conditions,
  like supporting TCP~\cite{Cerf74a,Postel80b,nitrd}.
Our new \emph{necessary} conditions
  determine
  when a network would \emph{leave} the Internet core.

\subsection{Defining the Internet Core}
	\label{sec:definition}

We define the Internet core as \emph{the strongly-connected component
  of more than 50\% of active, public IP addresses that can initiate communication with each other}.
Computers behind NATs and cloud load-balancers are on \emph{branches},
  participating but not part of the core, often with dynamically allocated,
  transient public IP addresses.
We believe this definition is simple, but with subtle implications.
For example, it defines \emph{two} Internet cores:
  one each for IPv4 and IPv6.

We build on the terms
  ``interconnected networks'', ``IP protocol'', and ``global address space''
  from prior definitions~\cite{Cerf74a,Postel80b,nitrd},
  and their common assumption
  that two computers on the public Internet
  should be able to reach each other directly
  at the IP layer.

We formalize ``network interconnection''~\cite{Cerf74a}
  by considering reachability over
  public IP addresses:
  addresses $x$ and $y$ are interconnected if traffic from $x$ can reach $y$
    and $y$ can reach $x$.
Networks are groups %
  mutually reachable addresses.

\textbf{Why more than 50\%?}
We take as an %
  axiom that there should be \emph{one Internet core} per address space,
  or a reason why that no core exists.
The definition must unambiguously identify ``the'' Internet core
  given conflicting claims.

Requiring a majority of active addresses
  ensures that there can be only one Internet core,
  since any two majorities must overlap
Any smaller fraction could allow two groups to make
  valid claims.
We discuss how to identify the core in the face of conflicting
  claims in \autoref{sec:half_proof}.

The definition of the Internet core should not require a central authority.
Majority supports assessment
  independent of any authority,
  as in other distributed consensus algorithms~\cite{Lamport82a,Lamport98a,Nakamoto09a}.
Any computer to prove it is in the Internet core
  by reaching half of active addresses,
  as defined by
  multiple, independent, long-term evaluations~\cite{Heidemann08c,Zander14b,Dainotti16a}.
We explicitly do not require identification of ``tier-1'' ISPs,
  an imprecise term often entangled with business concerns.

A majority defines an Internet core that can end:
  fragmentation occurs
  should the current Internet core break into three or more disconnected components
  where none retains a majority of active addresses.
If a large enough organization, nation, or group
  chose to secede, or are expelled,
  \emph{an} Internet core could become several no-longer internets.

\textbf{Why all and active addresses?}
In each of IPv4 and IPv6
  we consider all addresses equally. %
Public Internet addresses are global,
  and the Internet core intentionally designed without a hierarchy~\cite{clark1988design}.
Consistent with goals for
  network decentralization~\cite{dinrg},
  a definition should not create hierarchy,
  nor designate special addresses by age or importance.
\emph{Active} addresses are blocks that are reachable, defined below.

These definitions are relatively apolitical
  and reduce first-mover bias, discussed in \autoref{sec:internet_partition}.
Addresses are an Internet-centric metric,
  unlike population or countries.
Requiring activity
  reduces the influence of
  large allocated, but unused, space,
  such as in legacy IPv4 /8s
  and new IPv6 allocations.

\textbf{Reachability, Protocols and Firewalls:}
End-to-end reachability avoids
  difficult discovery of router-level topology.

Our conceptual definition allows different definitions of reachability.
Reachability can be tested by measurements with some protocol,
  such as ICMP echo-request (pings),
  or TCP or UDP queries,
  or by data-plane reachability with BGP\@.
Any specific test will provide an operational realization
  of our conceptual definition.
Particular tests will differ
  in how closely each approaches the conceptual ideal.

Firewalls complicate observing reachability,
  particularly when conditional or unidirectional.
We accept that the results of specific observations
  may vary with different protocols or observation times;
  practically we see results are stable
  with Internet-wide
  measurements~\cite{Baltra23a}.

We have two implementations of peninsula and island detection;
  both use publicly-available data from existing measurement systems.
One uses
  Trinocular~\cite{quan2013trinocular},
  because of its frequent, Internet-wide
  ICMP echo requests (11-minutes to 5M IPv4 /24s).
Prior work has shown ICMP provides the most response~\cite{Bartlett07d,quan2013trinocular,durumeric2014internet},
  and can avoid rate limiting~\cite{Guo18a},
  other other protocol options are possible.
Our second uses RIPE Atlas because of its use in DNS (\autoref{sec:dnsmon}).

\textbf{Why reachability and not applications?}
Users care about applications, and a user-centric view
  might emphasize reachability of HTTP or to Facebook
  rather than at the IP layer.
Our second realization uses public data from RIPE Atlas,
  with DNS as the application, as described
  in \autoref{sec:dnsmon}.
Future work may look at other, more user-centric applications.
However, we
  suggest reachability at the IP layer
  is a more fundamental concept.
IP has changed only twice since 1969 with IPv4 and IPv6,
  but dominant applications wax and wane,
  and applications such as e-mail
  extend beyond the Internet.

\textbf{Why strongly connected and bidirectional reachability?}
We require bidirectional reachability (strong connectivity)
  to identify \ac{NAT}-only computers as second class-citizens.
While most computers today are
  behind \ac{NAT} or cloud load-balancers,
  and NAT-ed computers are useful clients,
  they require protocols
  such as STUN~\cite{Rosenberg03a} to rendezvous through the core,
  or UPnP~\cite{Miller01a} or PMP~\cite{Cheshire13d} to link to the core.
Huge services run in the cloud by leasing public IP addresses from the cloud operator
  or importing their own (BYOIP).
Often services use a single public IP address
  but employ many servers
  behind a load balancer~\cite{Greenberg09a} or IP anycast~\cite{Partridge93a}.
While load balancers or home routers may be on the core,
  and some cloud VMs use fully-reachable public addresses,
  devices that are not bidirectionally reachable are not part of the core.

\subsection{Away from the Core: Unreachability}
\label{sec:internet_landscape}

\begin{figure}
  \begin{minipage}[b]{.56\columnwidth}
    \mbox{\includegraphics[width=\textwidth,trim=210 300 220 145,clip]{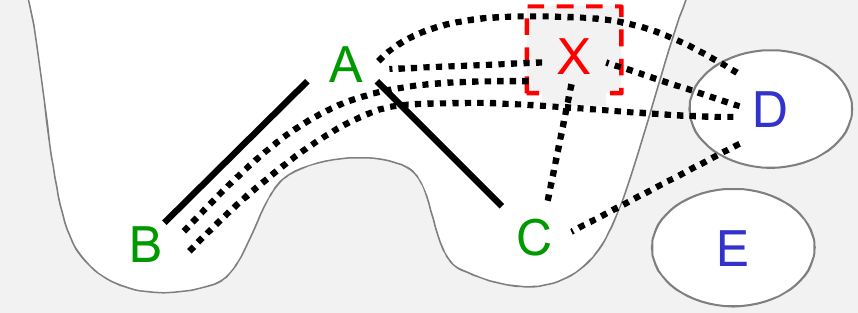}}
  \end{minipage}
  \begin{minipage}[b]{.42\columnwidth}
    \caption{Blocks $A$, $B$ and $C$ are the connected core,
      with $B$ and $C$ peninsulas;
      $D$ and $E$ islands;
      $X$ an outage.}
      \label{fig:term_concept}
  \end{minipage}
\end{figure}

We now use our definition of the Internet core to
  reason about where connectivity is incomplete:
  peninsulas, islands, and outages.
\autoref{fig:term_concept} shows
  a toy example,
  where long-term and current routability is shown by dotted and solid lines
  and white regions show current data-plane reachability.
All address blocks but $E$ form the core.
Blocks $B$ and $C$ are on \emph{peninsulas}
  because they do not route to each other,
  although data could relay through $A$.
Block $X$ has an \emph{outage};
  its routes are temporarily down.
Blocks $D$ and $E$ are \emph{islands}:
$D$ usually can route to the core,
  but not currently.
$E$ uses public addresses, but has never announced routes publicly.

\subsubsection{Outages}
\label{sec:outages}

A number of groups have examined Internet outages~\cite{Schulman11a,quan2013trinocular,richter2018advancing,guillot2019internet}.
These systems observe the public IPv4 Internet and identify networks
  that are no longer reachable---they have left the Internet.
Often these systems define outages operationally
  (network $X$ is out because none of our \acp{VP} can reach it).
In this paper, we define an outage as when all computers
  in a block are off,
  perhaps due to power loss.
We next define islands,
  when the computers are on but cannot reach the Internet core.

\subsubsection{Islands: Isolated Networks}
	\label{sec:island_definition}

An \emph{island} is a group of public IP addresses
  partitioned from the Internet core,
  but able to communicate among themselves.
Operationally, outages and islands are both unreachable from an external \ac{VP},
  but computers in an island are on and can reach each other.

Islands occur when an organization loses
  all connections to the Internet core.
A business with one office and one ISP becomes an island
  when its router's upstream connection fails,
  but computers in the office can reach each other and in-office servers.
An \emph{address island} is when
  a computer can ping only itself.
Externally, islands and outages appear identical.

\textbf{Example Islands:}
Islands are common in RIPE Atlas~\cite{Amin15a} when a \ac{VP}
  has an IPv6 address on the LAN, but lacks routes to the public IPv6 Internet.
In \autoref{sec:dnsmon} we show that this kind of misconfiguration
  accounts for
  $5\times$ more IPv6 unreachability than other, more meaningful problems.

We also see islands in reanalysis of data from Trinocular outage detection~\cite{quan2013trinocular}.
Over three years, from 2017 to 2020,
  we saw 14 cases where one of the 6 Trinocular \acp{VP}
  was active and could reach its LAN, but could not reach the rest of the Internet.
Network operators confirm local routing failures in several of these cases.

\subsubsection{Peninsulas: Partial Connectivity}
	\label{sec:peninsula_definition}

Link and power failures create islands,
  but \emph{partial} connectivity
  is  a more pernicious problem:
  when one can reach some destinations,
  but not others.
We call a group of public IP addresses with partial connectivity to the Internet core
  a \emph{peninsula}.
    In a geographic peninsula, the mainland may be visible over water, but reachable only with a detour; similarly,
    in \autoref{fig:term_concept},
    $B$ can reach $A$, but not $C$.
Peninsulas occur when
  an upstream provider of a multi-homed network
  accepts traffic but drops before delivery,
  when Tier-1 ISPs refuse to peer,
  or when firewalls block traffic.
Peninsula existence has long been recognized,
  prompting overlay networks to route around them~\cite{andersen2001resilient,katz2008studying,katz2012lifeguard}.

\textbf{Peninsulas in IPv6:}
An long-term peninsula follows from
  the IPv6 peering dispute between Hurricane Electric (HE) and Cogent.
These ISPs decline to peer in IPv6 (IPv4 is fine),
  nor do they forward their IPv6 through
  another party.
HE and Cogent customers could not reach each other in 2009~\cite{ipv6peeringdisputes},
  and this problem persists through 2024, as we show in DNSmon (\autoref{sec:dnsmon}).
We futher confirm unreachability
  between HE and Cogent users in IPv6 with traceroutes
  from looking glasses~\cite{he_looking_glass,cogent_looking_glass}
  (HE at 2001:470:20::2 and Cogent at 2001:550:1:a::d):
  neither can reach their neighbor's server,
  but both reach their own.

Other IPv6 disputes are Cogent with Google~\cite{google_cogent}, and
 Cloudflare with Hurricane Electric~\cite{cloudflare_he}.
Disputes are often due to an inability to  agree to %
  settlement-free or paid peering.

\textbf{Peninsulas in IPv4:}
We observed a peninsula lasting 3 hours starting 2017-10-23t22:02Z,
  where
  five Polish \acp{AS}
  had 1716 /24 blocks that were always reachable one Los Angeles,
  but not from four other \acp{VP}
  (as seen in public data from  Trinocular~\cite{LANDER17a}).
Before the peninsula, these blocks
  received service through Multimedia Polska (\emph{MP}, AS21021),
  via Cogent (AS174), or through Tata (AS6453).
When the peninsula occurred, traffic to all blocks continued through Cogent
  but was blackholed; it did not shift to Tata. %
The successful \ac{VP} could reach MP
  through  Tata for the entire event,
  proving MP was connected.
After 3 hours, we see a burst of 23k BGP updates
  and MP is again reachable from all VPs.
A graph showing reachability to this peninsula is in \autoref{sec:peninsula_exampla_data}.

We confirmed this peninsula with additional observations
  from traceroutes taken by CAIDA's Archipelago~\cite{CAIDA07b} (Ark).
During the event we see 94 unique Ark VPs attempted 345 traceroutes to the affected blocks.
Of the 94 VPs, 21 VPs (22\%) have their last responsive
  traceroute hop in the same \ac{AS} as the
  target address, and 68 (73\%) stopped before reaching that \ac{AS}.
The remaining 5 VPs were able to reach the destination \ac{AS} for only some
  traceroutes.
The large number of BGP updates suggest routing problems as a root cause.

\section{Applying the Definition}

\subsection{Resolving Conflicting Claims}
	\label{sec:half_proof}

Our definition
  of the Internet core in \autoref{sec:definition}
  must resolve conflicting claims
  without appeal to a central authority.

We can prove the definition yields a single core (or no core).
Consider a connected component with some fraction $A$, where $1 > A > 0.5$.
This component \emph{must} be larger than any other component $X$,
  as proven by contradiction:
  (i) assume some $X'$ exists, such that $X' > A$.
  (ii) Since $A > 0.5$, then (i) implies $X' > 0.5$.
  (iii) We then must conclude that $A+X' > 1$,
    but by definition, we measure only the whole address space,
    so it is also required that $A+X' \le 1$.
Therefore $X' < A$ and A forces a single clear component.
Q.E.D.

Disagree about what addresses are in the core
  can be resolved by comparing evidence.
Consider a simplified version of \autoref{fig:term_concept}
  with three pluralities
  of connectivity, $A$, $B$, and $C$, each representing one third of the addresses,
  where both $A$ and $B$  and $A$ and $C$ are strongly and directly connected,
  but $B$ and $C$ cannot directly reach each other.

In this example
  $A \cup B$ and $A \cup C$
  are partially, overlapping components of strong and direct connectivity,
  but since $B$ and $C$ cannot route to each other, they may dispute the core.
From our definition, \emph{all} ($A \cup B \cup C$)
  are in the core,
  but $B$ and $C$ are on peninsulas.
Any address can reach any other from either direction (the definition of ``strongly connected''),
  but since $B$ and $C$ do not exchange routes,
  they are partially connected peninsulas
  (unless one purchases transit from $A$).
These definitions apply if the sizes are about equal ($|A|=|B|=|C|=0.33$)
  or are asymmetric ($|A|=0.49$ and $|B|=|C|=0.02$).
Since the real Internet is mostly connected, typical values are
  $|A|>0.98$ and $|B|<0.01$.

Resolving competing claims require that all parties present
  their evidence (what addresses $A$, $B$, and $C$ can reach,
  and that they agree those addresses have the same meaning.
Private addresses and address squatting (described below)
  are cases where addresses have different meaning.

\subsection{Policy Applications of the Definition}
	\label{sec:policy_applications}

We next examine how a clear definition
  of the Internet core can inform policy tussles~\cite{Clark02a}.
Our hope is that our conceptual definition can make
  sometimes amorphous concepts like ``Internet fragmentation''
  more concrete,
  and an operational definition can quantify impacts
  and identify thresholds.

\textbf{Secession and Sovereignty:}
The U.S.~\cite{cybersecurity_act_2010}, China~\cite{Anonymous12a,Anonymous14a},
and Russia~\cite{russian_internet} have all proposed unplugging from
the Internet.
Egypt did in 2011~\cite{Cowie11a},
  and several countries have during exams~\cite{Gibbs16a,Dhaka18a,Henley18a,Economist18a}.
When the Internet partitions,
  which part is still ``the Internet core''?
Departure of an ISP or small country do not change the Internet core much,
  but what if a large country, or group of countries, leave together?

Our definition resolves this question, defining the Internet core
  from reachability of the majority of the active, public IP addresses
  (\autoref{sec:definition}).
Requiring a majority uniquely provides an unambiguous,
  externally evaluable test for the Internet core
  that allows one possible answer (the partition with more than 50\%).
In \autoref{sec:internet_partition} we discuss the corollary:
  the Internet core can end, turning into multiple partitions,
  if none retain a majority.
(A plurality is insufficient.)

\textbf{Sanction:}
An opposite of secession is expulsion.
Economic sanctions are one method of asserting international influence,
  and events such as the 2022 war in Ukraine prompted
  several large ISPs to discontinue service to Russia~\cite{Reuters22a}.
De-peering does not affect reachability for ISPs that purchase transit,
  but Tier-1 ISPs that de-peer create peninsulas for their users.
As described below in \autoref{sec:internet_partition},
  \emph{no single country can eject another by de-peering with it}.
However, a coalition of multiple countries could
  de-peer and eject a country from the Internet core
  if they, together, control
  more than half of the address space.

\subsection{Architecture and Operation Application}
	\label{sec:arch_applications}

Defining the core also helps clarify architectural changes such as the IPv4/v6 transition
  and operational %
  address reuse.

\textbf{The IPv4/v6 Transition:}
We have defined two Internet cores: IPv4 and IPv6.
Our definition can determine when one supersedes the other.
The networks will be on par when
  more than half of all IPv4 hosts are dual-homed.
After that point, IPv6 will supersede IPv4 when
  a majority of hosts on IPv6 can no longer reach IPv4.
Current limits on IPv6 measurement mean evaluation
  here is future work.
IPv6 shows the strength and limits of our definition:
  since IPv6 is already economically important,
  our definition seems irrelevant.
However, it may provide sharp boundary
  that makes the maturity of IPv6 definitive,
  helping motivate late-movers.

\textbf{Repurposing Addresses:}
Given full allocation of IPv4,
  multiple parties proposed re-purposing currently allocated or reserved IPv4 space,
  such 0/8 (``this'' network), 127/8 (loopback), and 240/4 (reserved)~\cite{Fuller08a}.
New use of these long-reserved addresses is challenged
  by assumptions in widely-deployed, difficult to change, existing software
  and hardware.
Our definition demonstrates
  that an RFC re-assigning this space for public traffic
  cannot make it a truly effective part of the Internet core until
  implementations used by a majority of active addresses
  can route to it.

\textbf{IPv4 Squat Space:}
IP squatting is when an organization
  requiring private address space beyond RFC1918
  takes over allocated but currently unrouted IPv4 space~\cite{Aronson15a}.
Several IPv4 /8s allocated to the U.S.~DoD have been used this way~\cite{Richter16c}
  (they were only publicly routed in 2021~\cite{Timberg21a}).
By our definition, such space is not part of the Internet core without
  public routes,
  and if more than half of the Internet is squatting on it,
  reclamation may be challenging.

\textbf{Internet outage detection:}
Outage detect systems of often reported confusing observations
  with a mix of positive and negative responses to active probes,
  such as ThunderPing's ``hosed'' state~\cite{Schulman11a}
  and observer-local problems in Trinocular~\cite{quan2013trinocular}.
Partial connectivity suggests that sometimes conflicting
  observations may be valid and peninsulas should be recognized
  legitimate occurrences for future exploration.

\textbf{Failure mitigation via routing:}
Several systems have proposed using relays to
  route around network-level routing failures~\cite{andersen2001resilient,katz2008studying,katz2012lifeguard}.
In addition, hypergiants operating their own backbones
  can select routing egress to avoid partial connectivity~\cite{Schlinker17a}.
An understanding of partial reachability in the WAN
  would quantify how important such efforts are.

\subsection{Can the Internet Core Partition?}
	\label{sec:internet_partition}

In \autoref{sec:policy_applications}
  we discussed secession and expulsion qualitatively.
Here we ask: Does any
  country or group have enough addresses to secede and claim to be
  ``the Internet core'' with a majority of addresses?
Alternatively,
  if a country were to exert control over their allocated addresses,
  would they become
  a country-sized island or peninsula?
We next use our reachability definition of more than 50\%
  to quantify control of the IP address space.

\begin{figure*}

\begin{minipage}[b]{.3\linewidth}
    \centering
    \resizebox{\textwidth}{!}{
    \begin{tabular}{l r r r r r r}
      & \multicolumn{4}{c}{\textbf{IPv4 Addresses}} &
      \multicolumn{2}{c}{\textbf{IPv6 Addresses}} \\
      \textbf{RIR}       &
      \multicolumn{2}{c}{\textbf{Active}}  &
      \multicolumn{2}{c}{\textbf{Allocated}}  &
      \multicolumn{2}{c}{\textbf{Allocated}} \\
      \midrule
      AFRINIC   & 15M    &  2\%   & 121M   & 3.3\%      & 9,661   & 3\%       \\
      APNIC     & 223M   & \cellcolor[HTML]{99ee77}33\%   & 892M  & 24.0\%      & 88,614  & 27.8\%    \\
      \rowcolor[HTML]{DCDCDC}
      \hspace{1mm} \emph{China}   & 112M & 17\% &  345M   &    9.3\%     & 54,849  & \cellcolor[HTML]{FFF9C4}17.2\%    \\
      ARIN      & 150M   & 22\%   & 1,673M & \cellcolor[HTML]{99ee77}45.2\%      & 56,172  & 17.6\%    \\
      \rowcolor[HTML]{DCDCDC}
      \hspace{1mm} \emph{U.S.}    & 140M & \cellcolor[HTML]{FFF9C4}21\% & 1,617M  & \cellcolor[HTML]{FFF9C4}43.7\%        & 55,026  & \cellcolor[HTML]{FFF9C4}17.3\%  \\
      LACNIC    & 82M  & 12\%     & 191M  & 5.2\%      & 15,298  & 4.8\%     \\
      RIPE NCC  & 206M & 30\%     & 826M  & 22.3\%      & 148,881 & \cellcolor[HTML]{99ee77}46.7\%    \\
      \rowcolor[HTML]{DCDCDC}
      \hspace{1mm} \emph{Germany} & 40M & 6\% & 124M  &    3.3\%     & 22,075  & 6.9\%     \\
      \midrule
      Total &  676M &  100\% & 3,703M & 100\%      & 318,626 & 100\%     \\
    \end{tabular}}
    \captionsetup{type=table}
    \caption{RIR IPv4 hosts and IPv6 /32 alloc.  \cite{iana_v4, iana_v6}. %
     }
    \label{tab:rir_allocation}
\end{minipage}
\hspace{2mm}
\begin{minipage}[b]{.31\linewidth}
    \centering
    \footnotesize
    \resizebox{1\textwidth}{!}{
    \includegraphics[trim=0 0 0 5,clip,width=1\textwidth]{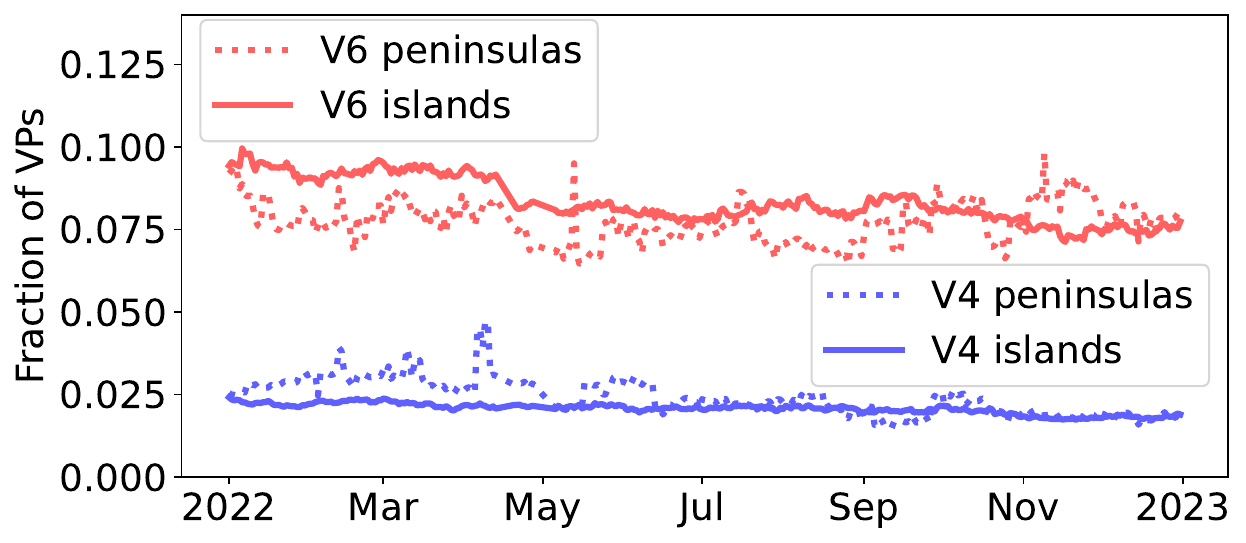}}
        \vspace*{-4mm}
    \captionsetup{type=figure}
    \captionsetup{width=0.9\textwidth}
    \captionof{figure}{Fraction of VPs observing islands and peninsulas for IPv4 and IPv6 during 2022.}
  \label{fig:a49_partial_outages}
\end{minipage}
\hspace{2mm}
\begin{minipage}[b]{.31\linewidth}
    \centering
    	\footnotesize
        \resizebox{1\textwidth}{!}{
        \includegraphics[trim=0 0 0 5,clip,width=1\textwidth]{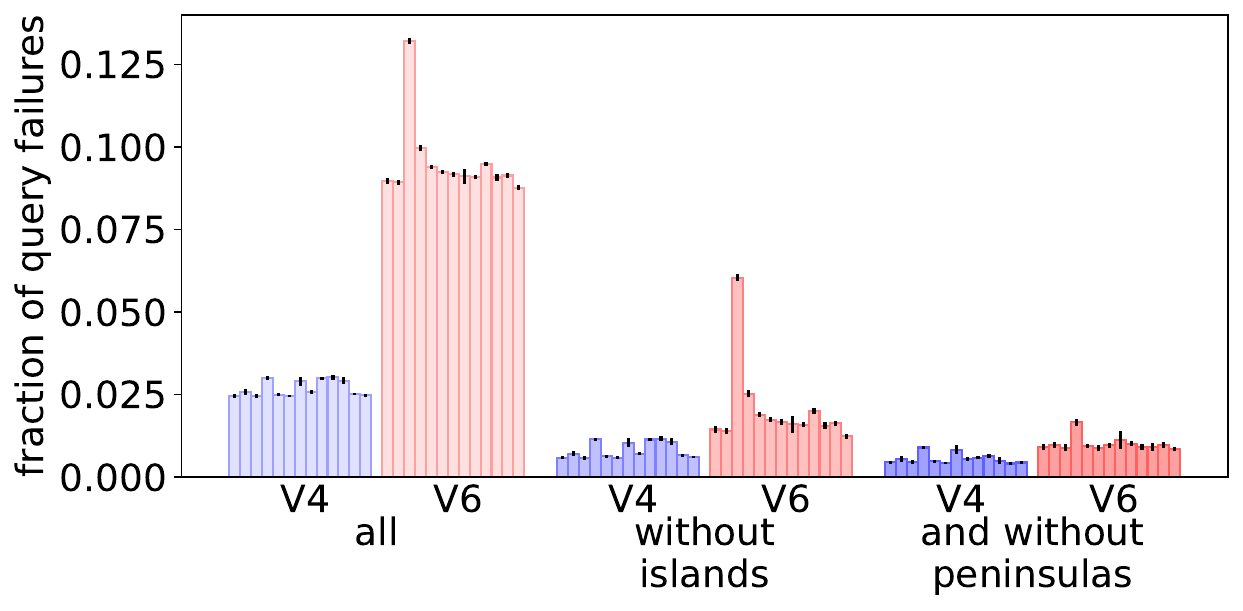}}
        \vspace*{-6mm}
    \captionsetup{type=figure}
    \captionof{figure}{Atlas queries from all available VPs to 13
    Root Servers for IPv4 and IPv6 on 2022-07-23.}
    \label{fig:atlas_revisited}
\end{minipage}
\end{figure*}

To evaluate the power of countries and \acp{RIR} over the Internet core,
  \autoref{tab:rir_allocation} reports the number of active IPv4
  addresses as determined by Internet censuses~\cite{Heidemann08c}
  for \acp{RIR} and selected countries.
Since estimating active IPv6 addresses is an open problem,
  we provide allocated addresses for both v4 and v6~\cite{iana_v4,
  iana_v6}.
(IPv4 has been fully allocated since 2011~\cite{ICANN11a}).

\autoref{tab:rir_allocation} shows that \emph{no individual \ac{RIR} or country can secede and take the Internet core},
  because none controls the majority of IPv4 addresses.
ARIN has the largest share with 1673M allocated (45.2\%).
Of countries, U.S. has the largest share of allocated IPv4 (1617M, 43.7\%).
Active addresses are more evenly distributed
  with APNIC (223M, 33\%) and the U.S.~(40M, 21\%) the largest \ac{RIR} and country.

\emph{IPv6 is also an international collaboration},
  since no \ac{RIR} or country is exceeds 50\% allocation.
RIPE (an RIR) is close with 46.7\%,
  and China and the U.S.~have large country allocations.
With most of IPv6 unallocated, these fractions may change.

IPv4 reflects a first-mover bias, where early adopters acquired
  many addresses,
  but this factor is smaller in IPv6.
Our definition's use of active addresses also reduces this bias,
  since
  numbers of \emph{active} IPv4 addresses
  is similar to allocated IPv6 addresses
  (legacy IPv4 addresses are less used).

\subsection{Definitions Clarify DNSmon Sensitivity}
	\label{sec:dnsmon}

We next show how understanding partial connectivity can
  improve DNSmon sensitivity.
DNSmon~\cite{Amin15a}
  monitors the Root Server System~\cite{RootServers16a}
  from about 10k RIPE Atlas VPs (probes)~\cite{Ripe15c}.
For years, DNSmon has reported IPv6 loss rates of 4-10\%,
  $4\times$ higher than IPv4.
The DNS root is well provisioned and distributed,
  so why is IPv6 loss so high?

RIPE Atlas operators are aware of problems with some Atlas VPs.
Some VPs support IPv6 on their LAN, but not to the global IPv6 Internet---such VPs
  are IPv6 islands.
Atlas periodically tags and culls these VPs
  from DNSmon.
However, our study of DNSmon
  for islands and peninsulas
  improves their results.
Using concepts pioneered here (\autoref{sec:problem}),
  we give full analysis in a
    workshop paper~\cite{Saluja22a};
Here we add new data
  showing these results persist for 1 year (\autoref{fig:a49_partial_outages}).

Groups of bars in
  \autoref{fig:atlas_revisited} show query loss
  for each of the 13 root service identifiers,
  as observed
  from all available Atlas VPs (10,082 IPv4, and 5,173 IPv6)
  on 2022-07-23.
(We are similar to DNSmon, but it uses only about 100 well-connected ``anchors'',
  so our analysis is wider.)
The first two groups show loss rates for IPv4 (light blue, left most) and IPv6 (light red),
  showing IPv4 losses around 2\%, and IPv6 from 9 to 13\%.

We report a VP as an island when it
  cannot see \emph{any} of the 13 root identifiers over 24~hours.
(This definition is stricter than our 50\% definition
  since VPs attempt only 13 targets, not the whole Internet,
  and we apply it over a full day to consider only long-term trends.)
The middle two groups of bars show IPv4 and IPv6 loss rates
  after removing VPs that
  are islands.
Without island VPs,
  IPv4 loss rates drop to 0.005 from 0.01, and IPv6 to about 0.01 from 0.06.
These rates represent a more
  meaningful estimate of DNS reliability.
Users of VPs that are IPv6 islands
  will not expect global IPv6,
  and such VPs should not be used for IPv6 in DNSmon.

The third bar in each red cluster of IPv6 is an outlier:
  that root identifier shows 13\% IPv6 loss with all VPs,
  and 6\% loss after islands are removed.
This result is explained by
  persistent routing disputes between Cogent (the operator of C-Root) and Hurricane Electric~\cite{Miller09a}.
Omitting islands (the middle bars) makes this difference much clearer.

Finally we detect peninsulas by looking for \acp{VP} that
  each some but not all root servers.
Peninsulas suggest persistent routing problems;
  they deserve attention from ISPs and root operators.
The darker, rightmost two groups show loss from non-island/peninsula VPs,
  representing loss if routing problems were addressed.
With this correction C-Root is similar to others,
  confirming peering disputes affect its success.

This example shows how \emph{understanding partial reachability
  can improve the sensitivity of existing measurement systems}.
Removing islands makes it easy to identify persistent routing problems.
Removing peninsulas makes
  transient changes (perhaps from failure, DDoS, routing)
  more visible.
Each layer of problem is important,
  but by considering each separately,
  the interesting ``signal'' of routing changes
  (appearing in the right two groups in \autoref{fig:atlas_revisited}),
  appears out from under the
  $5\times$ or $9.7\times$ times larger peninsulas and islands (the left two groups).
Improved sensitivity also \emph{shows a need to improve
  IPv6 provisioning},
  since
  IPv6 loss is statistically higher than
  IPv4 loss (compare the right blue and red groups),
  even accounting for known problems.
After sharing the results with root operators and RIPE Atlas,
  two operators adopted them in regular operation.

\section{Related Work}
	\label{sec:related}

Prior definitions of the
    Internet exist at the IP-layer~\cite{Cerf74a,Postel80b,nitrd,huawei2020}
    of their time, or the AS-level~\cite{Gao01b,Luckie13a}.
IPNL proposed a core-only Internet with all users behind NAT~\cite{Francis01b}.
We instead consider the IP-layer in today's architecture
  to address today's challenges (see \autoref{sec:problem}).

Several systems mitigate partial outages.
RON provides alternate-path routing
  around failures for a mesh of sites~\cite{andersen2001resilient}.
Hubble monitors multi-path reachability~\cite{katz2008studying}.
LIFEGUARD routes around reachability failures~\cite{katz2012lifeguard}.
These systems address partial reachability;
  we define its scope.

Prior work studied partial reachability, showing
  it is a common transient occurrence
  during routing convergence~\cite{bush2009internet}.
They reproduced partial connectivity with controlled experiments;
  we study it in RIPE Atlas.

Active outage detection systems have encountered partial outages.
ThunderPing recognizes a ``hosed'' state with mixed replies,
  but its study is future work~\cite{Schulman11a}.
Trinocular discards partial outages by
  reporting the target block ``up'' if any VP can reach
  it~\cite{quan2013trinocular}.
To the best of our knowledge, prior outage detection systems
  do not consistently report partial outages in the Internet core,
  nor do they study their extent.

We use the idea of majority to define the Internet core in the face of secession.
That idea is fundamental in many algorithms for distributed consensus~\cite{Lamport82a,Lamport98a,Nakamoto09a},
  with applications for example to certificate authorities~\cite{birge2018bamboozling}.

Recent groups have studied the policy issues around Internet fragmentation~\cite{Drake16a,Drake22a}, but do not define it.
We hope our definition can fill that need.

\section{Conclusions}

This paper affirms the importance of an Internet core for
  global communication,
  and provides a robust, operationalizable definition of that core.
The definition helps identify
  disconnected islands and shows that partially connected peninsulas
  are an important challenge.
The definition helps clarify what events would cause the Internet core
  to fragment.
They also help help improve the sensitivity of
  operational measurement systems such as
  RIPE DNSmon, by distinguishing long-term partial reachability
  from short-term changes.

\begin{acks}

The authors would like to thank John Wroclawski,
  Wes Hardaker, Ramakrishna Padmanabhan,
  Ramesh Govindan,
  Eddie Kohler,
  and the Internet Architecture Board for their input on
  on an early version of this paper.

The work is
  supported in part by
   the National Science Foundation, CISE
  Directorate, award CNS-2007106 %
  and NSF-2028279. %
The U.S.~Government is authorized to reproduce and distribute
reprints for Governmental purposes notwithstanding any copyright
notation thereon.
\end{acks}

\label{page:last_body}

\bibliographystyle{ACM-Reference-Format}


\begin{thebibliography}{81}


\ifx \showCODEN    \undefined \def \showCODEN     #1{\unskip}     \fi
\ifx \showDOI      \undefined \def \showDOI       #1{#1}\fi
\ifx \showISBNx    \undefined \def \showISBNx     #1{\unskip}     \fi
\ifx \showISBNxiii \undefined \def \showISBNxiii  #1{\unskip}     \fi
\ifx \showISSN     \undefined \def \showISSN      #1{\unskip}     \fi
\ifx \showLCCN     \undefined \def \showLCCN      #1{\unskip}     \fi
\ifx \shownote     \undefined \def \shownote      #1{#1}          \fi
\ifx \showarticletitle \undefined \def \showarticletitle #1{#1}   \fi
\ifx \showURL      \undefined \def \showURL       {\relax}        \fi
\providecommand\bibfield[2]{#2}
\providecommand\bibinfo[2]{#2}
\providecommand\natexlab[1]{#1}
\providecommand\showeprint[2][]{arXiv:#2}

\bibitem[\protect\citeauthoryear{Amin, C{\'a}ndela, Karrenberg, Kisteleki, and
  Strikos}{Amin et~al\mbox{.}}{2015}]%
        {Amin15a}
\bibfield{author}{\bibinfo{person}{Christopher Amin}, \bibinfo{person}{Massimo
  C{\'a}ndela}, \bibinfo{person}{Daniel Karrenberg}, \bibinfo{person}{Robert
  Kisteleki}, {and} \bibinfo{person}{Andreas Strikos}.}
  \bibinfo{year}{2015}\natexlab{}.
\newblock \showarticletitle{Visualization and Monitoring for the Identification
  and Analysis of {DNS} Issues}. In \bibinfo{booktitle}{\emph{Proceedings of
  the International Conference on the Internet Monitoring and Protection}}.
  \bibinfo{address}{Brussels, Belgium}.
\newblock
\urldef\tempurl%
\url{https://www.researchgate.net/profile/Massimo-Candela/publication/279516870_Visualization_and_Monitoring_for_the_Identification_and_Analysis_of_DNS_Issues/links/559468c808ae793d13798901/Visualization-and-Monitoring-for-the-Identification-and-Analysis-of-DNS-Issues.pdf}
\showURL{%
\tempurl}


\bibitem[\protect\citeauthoryear{Andersen, Balakrishnan, Kaashoek, and
  Morris}{Andersen et~al\mbox{.}}{2001}]%
        {andersen2001resilient}
\bibfield{author}{\bibinfo{person}{David~G. Andersen}, \bibinfo{person}{Hari
  Balakrishnan}, \bibinfo{person}{M.~Frans Kaashoek}, {and}
  \bibinfo{person}{Robert Morris}.} \bibinfo{year}{2001}\natexlab{}.
\newblock \showarticletitle{Resilient Overlay Networks}. In
  \bibinfo{booktitle}{\emph{Proceedings of the Symposium on Operating Systems
  Principles}}. \bibinfo{publisher}{{ACM}}, \bibinfo{address}{Chateau Lake
  Louise, Alberta, Canada}, \bibinfo{pages}{131--145}.
\newblock
\urldef\tempurl%
\url{http://www-cse.ucsd.edu/sosp01/papers/andersen.pdf}
\showURL{%
\tempurl}


\bibitem[\protect\citeauthoryear{Anonymous}{Anonymous}{2012}]%
        {Anonymous12a}
\bibfield{author}{\bibinfo{person}{Anonymous}.}
  \bibinfo{year}{2012}\natexlab{}.
\newblock \showarticletitle{The collateral damage of {Internet} censorship by
  {DNS} injection}.
\newblock \bibinfo{journal}{\emph{{ACM} Computer Communication Review}}
  \bibinfo{volume}{42}, \bibinfo{number}{3} (\bibinfo{date}{July}
  \bibinfo{year}{2012}), \bibinfo{pages}{21--27}.
\newblock
\urldef\tempurl%
\url{https://doi.org/10.1145/2317307.2317311}
\showDOI{\tempurl}


\bibitem[\protect\citeauthoryear{Anonymous}{Anonymous}{2014}]%
        {Anonymous14a}
\bibfield{author}{\bibinfo{person}{Anonymous}.}
  \bibinfo{year}{2014}\natexlab{}.
\newblock \showarticletitle{Towards a Comprehensive Picture of the {Great}
  {Firewall's} {DNS} Censorship}. In \bibinfo{booktitle}{\emph{Proceedings of
  the {USENIX} Workshop on Free and Open Communciations on the Internet}}.
  \bibinfo{publisher}{{USENIX}}, \bibinfo{address}{San Diego, CA, USA}, 7.
\newblock
\urldef\tempurl%
\url{https://www.usenix.org/system/files/conference/foci14/foci14-anonymous.pdf}
\showURL{%
\tempurl}


\bibitem[\protect\citeauthoryear{{ANT Project}}{{ANT Project}}{2022}]%
        {ANT22b}
\bibfield{author}{\bibinfo{person}{{ANT Project}}.}
  \bibinfo{year}{2022}\natexlab{}.
\newblock \bibinfo{title}{{ANT} {IPv4} Island and Peninsula Data}.
\newblock
  \bibinfo{howpublished}{\url{https://ant.isi.edu/datasets/ipv4_partial/}}.
\newblock
\urldef\tempurl%
\url{https://ant.isi.edu/datasets/ipv4_partial/}
\showURL{%
\tempurl}


\bibitem[\protect\citeauthoryear{Aronson}{Aronson}{2015}]%
        {Aronson15a}
\bibfield{author}{\bibinfo{person}{Cathy Aronson}.}
  \bibinfo{year}{2015}\natexlab{}.
\newblock \bibinfo{title}{To Squat Or Not To Squat?}
\newblock \bibinfo{howpublished}{blog
  \url{https://teamarin.net/2015/11/23/to-squat-or-not-to-squat/}}.
\newblock
\urldef\tempurl%
\url{https://teamarin.net/2015/11/23/to-squat-or-not-to-squat/}
\showURL{%
\tempurl}


\bibitem[\protect\citeauthoryear{Baltra and Heidemann}{Baltra and
  Heidemann}{2020}]%
        {Baltra20a}
\bibfield{author}{\bibinfo{person}{Guillermo Baltra} {and}
  \bibinfo{person}{John Heidemann}.} \bibinfo{year}{2020}\natexlab{}.
\newblock \showarticletitle{Improving Coverage of Internet Outage Detection in
  Sparse Blocks}. In \bibinfo{booktitle}{\emph{Proceedings of the Passive and
  Active Measurement Workshop}}. \bibinfo{publisher}{Springer},
  \bibinfo{address}{Eugene, Oregon, USA}.
\newblock


\bibitem[\protect\citeauthoryear{Baltra and Heidemann}{Baltra and
  Heidemann}{2023}]%
        {Baltra23a}
\bibfield{author}{\bibinfo{person}{Guillermo Baltra} {and}
  \bibinfo{person}{John Heidemann}.} \bibinfo{year}{2023}\natexlab{}.
\newblock \bibinfo{booktitle}{\emph{Detecting Partial Reachability in the
  {Internet} Core}}.
\newblock \bibinfo{type}{{T}echnical {R}eport} arXiv:2107.11439v4.
  \bibinfo{institution}{USC/Information Sciences Institute}.
\newblock
\urldef\tempurl%
\url{https://doi.org/10.48550/2107.11439v4}
\showDOI{\tempurl}


\bibitem[\protect\citeauthoryear{Bartlett, Heidemann, and
  Papadopoulos}{Bartlett et~al\mbox{.}}{2007}]%
        {Bartlett07d}
\bibfield{author}{\bibinfo{person}{Genevieve Bartlett}, \bibinfo{person}{John
  Heidemann}, {and} \bibinfo{person}{Christos Papadopoulos}.}
  \bibinfo{year}{2007}\natexlab{}.
\newblock \showarticletitle{Understanding Passive and Active Service
  Discovery}. In \bibinfo{booktitle}{\emph{Proceedings of the ACM Internet
  Measurement Conference}}. \bibinfo{publisher}{{ACM}}, \bibinfo{address}{San
  Diego, California, USA}, \bibinfo{pages}{57--70}.
\newblock
\urldef\tempurl%
\url{https://doi.org/10.1145/1298306.1298314}
\showDOI{\tempurl}


\bibitem[\protect\citeauthoryear{Birge-Lee, Sun, Edmundson, Rexford, and
  Mittal}{Birge-Lee et~al\mbox{.}}{2018}]%
        {birge2018bamboozling}
\bibfield{author}{\bibinfo{person}{Henry Birge-Lee}, \bibinfo{person}{Yixin
  Sun}, \bibinfo{person}{Anne Edmundson}, \bibinfo{person}{Jennifer Rexford},
  {and} \bibinfo{person}{Prateek Mittal}.} \bibinfo{year}{2018}\natexlab{}.
\newblock \showarticletitle{Bamboozling certificate authorities with {BGP}}. In
  \bibinfo{booktitle}{\emph{27th USENIX Security Symposium}}.
  \bibinfo{publisher}{{USENIX}}, \bibinfo{address}{Baltimore, Maryland, USA},
  \bibinfo{pages}{833--849}.
\newblock


\bibitem[\protect\citeauthoryear{Bush, Maennel, Roughan, and Uhlig}{Bush
  et~al\mbox{.}}{2009}]%
        {bush2009internet}
\bibfield{author}{\bibinfo{person}{Randy Bush}, \bibinfo{person}{Olaf Maennel},
  \bibinfo{person}{Matthew Roughan}, {and} \bibinfo{person}{Steve Uhlig}.}
  \bibinfo{year}{2009}\natexlab{}.
\newblock \showarticletitle{Internet optometry: assessing the broken glasses in
  Internet reachability}. In \bibinfo{booktitle}{\emph{Proceedings of the 9th
  ACM SIGCOMM conference on Internet measurement}}. \bibinfo{publisher}{{ACM}},
  \bibinfo{address}{Chicago, Illinois, USA}, \bibinfo{pages}{242--253}.
\newblock
\urldef\tempurl%
\url{http://www.maennel.net/2009/imc099-bush.pdf}
\showURL{%
\tempurl}


\bibitem[\protect\citeauthoryear{{CAIDA}}{{CAIDA}}{2007}]%
        {CAIDA07b}
\bibfield{author}{\bibinfo{person}{{CAIDA}}.} \bibinfo{year}{2007}\natexlab{}.
\newblock \bibinfo{title}{Archipelago (Ark) Measurement Infrastructure}.
\newblock \bibinfo{howpublished}{website
  \url{https://www.caida.org/projects/ark/}}.
\newblock
\urldef\tempurl%
\url{https://www.caida.org/projects/ark/}
\showURL{%
\tempurl}


\bibitem[\protect\citeauthoryear{Cerf and Kahn}{Cerf and Kahn}{1974}]%
        {Cerf74a}
\bibfield{author}{\bibinfo{person}{Vint Cerf} {and} \bibinfo{person}{Robert
  Kahn}.} \bibinfo{year}{1974}\natexlab{}.
\newblock \showarticletitle{A Protocol for Packet Network Interconnection}.
\newblock \bibinfo{journal}{\emph{IEEE Transactions on Communications}}
  \bibinfo{volume}{COM-22}, \bibinfo{number}{5} (\bibinfo{date}{May}
  \bibinfo{year}{1974}), \bibinfo{pages}{637--648}.
\newblock
\urldef\tempurl%
\url{http://sysnet.ucsd.edu/classes/cse222/wi03/papers/cerf-tcp-toc74.pdf}
\showURL{%
\tempurl}


\bibitem[\protect\citeauthoryear{Cheshire and Krochmal}{Cheshire and
  Krochmal}{2013}]%
        {Cheshire13d}
\bibfield{author}{\bibinfo{person}{S. Cheshire} {and} \bibinfo{person}{M.
  Krochmal}.} \bibinfo{year}{2013}\natexlab{}.
\newblock \bibinfo{booktitle}{\emph{{NAT} Port Mapping Protocol ({NAT-PMP})}}.
\newblock \bibinfo{type}{RFC} 6886. \bibinfo{institution}{Internet Request For
  Comments}.
\newblock
\urldef\tempurl%
\url{https://doi.org/10.17487/RFC6886}
\showDOI{\tempurl}


\bibitem[\protect\citeauthoryear{Clark}{Clark}{1988}]%
        {clark1988design}
\bibfield{author}{\bibinfo{person}{David~D. Clark}.}
  \bibinfo{year}{1988}\natexlab{}.
\newblock \showarticletitle{The Design Philosophy of the {DARPA} {Internet}
  Protocols}. In \bibinfo{booktitle}{\emph{Proceedings of the 1988 Symposium on
  Communications Architectures and Protocols}}. \bibinfo{publisher}{{ACM}},
  \bibinfo{pages}{106--114}.
\newblock


\bibitem[\protect\citeauthoryear{Clark, Wroclawski, Sollins, and Braden}{Clark
  et~al\mbox{.}}{2002}]%
        {Clark02a}
\bibfield{author}{\bibinfo{person}{David~D. Clark}, \bibinfo{person}{John
  Wroclawski}, \bibinfo{person}{Karen Sollins}, {and} \bibinfo{person}{Robert
  Braden}.} \bibinfo{year}{2002}\natexlab{}.
\newblock \showarticletitle{Tussle in Cyberspace: Defining Tomorrow's
  Internet}. In \bibinfo{booktitle}{\emph{Proceedings of the {ACM} SIGCOMM
  Conference}}. \bibinfo{publisher}{{ACM}}, \bibinfo{address}{Pittsburgh, PA,
  USA}, \bibinfo{pages}{347--356}.
\newblock
\urldef\tempurl%
\url{http://www.acm.org/sigcomm/sigcomm2002/papers/tussle.pdf}
\showURL{%
\tempurl}


\bibitem[\protect\citeauthoryear{{CNBC}}{{CNBC}}{2019}]%
        {russian_internet}
\bibfield{author}{\bibinfo{person}{{CNBC}}.} \bibinfo{year}{2019}\natexlab{}.
\newblock \bibinfo{title}{Russia just brought in a law to try to disconnect its
  {Internet} from the rest of the world}.
\newblock
  \bibinfo{howpublished}{\url{https://www.cnbc.com/2019/11/01/russia-controversial-sovereign-internet-law-goes-into-force.html}}.
\newblock


\bibitem[\protect\citeauthoryear{Coca}{Coca}{2018}]%
        {Coca18a}
\bibfield{author}{\bibinfo{person}{N. Coca}.} \bibinfo{year}{2018}\natexlab{}.
\newblock \showarticletitle{China's Xinjiang surveillance is the dystopian
  future nobody wants}.
\newblock \bibinfo{journal}{\emph{Engadget}} (\bibinfo{date}{Feb.~22}
  \bibinfo{year}{2018}).
\newblock
\urldef\tempurl%
\url{https://www.engadget.com/2018-02-22-china-xinjiang-surveillance-tech-spread.html}
\showURL{%
\tempurl}


\bibitem[\protect\citeauthoryear{Cogent}{Cogent}{2021}]%
        {cogent_looking_glass}
\bibfield{author}{\bibinfo{person}{Cogent}.} \bibinfo{year}{2021}\natexlab{}.
\newblock \bibinfo{title}{Looking Glass}.
\newblock \bibinfo{howpublished}{\url{https://cogentco.com/en/looking-glass}}.
\newblock


\bibitem[\protect\citeauthoryear{Cowie}{Cowie}{2011}]%
        {Cowie11a}
\bibfield{author}{\bibinfo{person}{James Cowie}.}
  \bibinfo{year}{2011}\natexlab{}.
\newblock \bibinfo{title}{{Egypt} Leaves the {Internet}}.
\newblock \bibinfo{howpublished}{Renesys Blog
  \url{http://www.renesys.com/blog/2011/01/egypt-leaves-the-internet.shtml}}.
\newblock
\urldef\tempurl%
\url{http://www.renesys.com/blog/2011/01/egypt-leaves-the-internet.shtml}
\showURL{%
\tempurl}


\bibitem[\protect\citeauthoryear{daily}{daily}{2021}]%
        {RBC21a}
\bibfield{author}{\bibinfo{person}{RBC daily}.}
  \bibinfo{year}{2021}\natexlab{}.
\newblock \bibinfo{title}{Russia, tested the Runet when disconnected from the
  Global Network}.
\newblock \bibinfo{howpublished}{website
  \url{https://www.rbc.ru/technology_and_media/21/07/2021/60f8134c9a79476f5de1d739}}.
\newblock
\urldef\tempurl%
\url{https://www.rbc.ru/technology_and_media/21/07/2021/60f8134c9a79476f5de1d739}
\showURL{%
\tempurl}


\bibitem[\protect\citeauthoryear{Dainotti, Benson, King, kc~claffy, Glatz,
  Dimitropoulos, Richter, Finamore, and Snoeren}{Dainotti
  et~al\mbox{.}}{2016}]%
        {Dainotti16a}
\bibfield{author}{\bibinfo{person}{Alberto Dainotti}, \bibinfo{person}{Karyn
  Benson}, \bibinfo{person}{Alistair King}, \bibinfo{person}{kc claffy},
  \bibinfo{person}{Eduard Glatz}, \bibinfo{person}{Xenofontas Dimitropoulos},
  \bibinfo{person}{Philipp Richter}, \bibinfo{person}{Alessandro Finamore},
  {and} \bibinfo{person}{Alex~C. Snoeren}.} \bibinfo{year}{2016}\natexlab{}.
\newblock \showarticletitle{Lost in Space: Improving Inference of {IPv4}
  Address Space Utilization}.
\newblock \bibinfo{journal}{\emph{{IEEE} Journal of Selected Areas in
  Communication}} \bibinfo{volume}{34}, \bibinfo{number}{6}
  (\bibinfo{date}{April} \bibinfo{year}{2016}), \bibinfo{pages}{1862--1876}.
\newblock
\urldef\tempurl%
\url{https://doi.org/10.1109/JSAC.2016.2559218}
\showDOI{\tempurl}


\bibitem[\protect\citeauthoryear{{Dhaka Tribune Desk}}{{Dhaka Tribune
  Desk}}{2018}]%
        {Dhaka18a}
\bibfield{author}{\bibinfo{person}{{Dhaka Tribune Desk}}.}
  \bibinfo{year}{2018}\natexlab{}.
\newblock \showarticletitle{{Internet} services to be suspended across the
  country}.
\newblock \bibinfo{journal}{\emph{Dhaka Tribune}} (\bibinfo{date}{Feb. 11}
  \bibinfo{year}{2018}).
\newblock
\urldef\tempurl%
\url{http://www.dhakatribune.com/regulation/2018/02/11/internet-services-suspended-throughout-country/}
\showURL{%
\tempurl}


\bibitem[\protect\citeauthoryear{Dhamdhere, Clark, Gamero-Garrido, Luckie, Mok,
  Akiwate, Gogia, Bajpai, Snoeren, and kc~claffy}{Dhamdhere
  et~al\mbox{.}}{2018}]%
        {Dhamdhere18a}
\bibfield{author}{\bibinfo{person}{Amogh Dhamdhere}, \bibinfo{person}{David~D.
  Clark}, \bibinfo{person}{Alexander Gamero-Garrido}, \bibinfo{person}{Matthew
  Luckie}, \bibinfo{person}{Ricky K.~P. Mok}, \bibinfo{person}{Gautam Akiwate},
  \bibinfo{person}{Kabir Gogia}, \bibinfo{person}{Vaibhav Bajpai},
  \bibinfo{person}{Alex~C. Snoeren}, {and} \bibinfo{person}{kc claffy}.}
  \bibinfo{year}{2018}\natexlab{}.
\newblock \showarticletitle{Inferring Persistent Interdomain Congestion}. In
  \bibinfo{booktitle}{\emph{Proceedings of the {ACM} SIGCOMM Conference}}.
  \bibinfo{publisher}{{ACM}}, \bibinfo{address}{Budapest, Hungary},
  \bibinfo{pages}{1--15}.
\newblock
\urldef\tempurl%
\url{https://doi.org/10.1145/3230543.3230549}
\showDOI{\tempurl}


\bibitem[\protect\citeauthoryear{{DINRG}}{{DINRG}}{2021}]%
        {dinrg}
\bibfield{author}{\bibinfo{person}{{DINRG}}.} \bibinfo{year}{2021}\natexlab{}.
\newblock \bibinfo{title}{Decentralized Internet Infrastructure Research
  Group}.
\newblock \bibinfo{howpublished}{\url{https://irtf.org/dinrg}}.
\newblock


\bibitem[\protect\citeauthoryear{Drake, Cerf, and Kleinw{\"a}chter}{Drake
  et~al\mbox{.}}{2016}]%
        {Drake16a}
\bibfield{author}{\bibinfo{person}{William~J. Drake},
  \bibinfo{person}{Vinton~G. Cerf}, {and} \bibinfo{person}{Wolfgang
  Kleinw{\"a}chter}.} \bibinfo{year}{2016}\natexlab{}.
\newblock \bibinfo{booktitle}{\emph{Internet Fragmentation: An Overview}}.
\newblock \bibinfo{type}{{T}echnical {R}eport}. \bibinfo{institution}{World
  Economic Forum}.
\newblock
\urldef\tempurl%
\url{https://www3.weforum.org/docs/WEF_FII_Internet_Fragmentation_An_Overview_2016.pdf}
\showURL{%
\tempurl}


\bibitem[\protect\citeauthoryear{Drake~(moderator)}{Drake~(moderator)}{2022}]%
        {Drake22a}
\bibfield{author}{\bibinfo{person}{William~J. Drake~(moderator)}.}
  \bibinfo{year}{2022}\natexlab{}.
\newblock \bibinfo{title}{Internet Fragmentation, Reconsidered}.
\newblock \bibinfo{howpublished}{CITI Seminar on Global Digital Governance at
  IETF 115}.
\newblock
\urldef\tempurl%
\url{https://www8.gsb.columbia.edu/citi/GlobalDigitalGovernance}
\showURL{%
\tempurl}


\bibitem[\protect\citeauthoryear{Dunn}{Dunn}{2021}]%
        {scientific_methods}
\bibfield{author}{\bibinfo{person}{Peter~K. Dunn}.}
  \bibinfo{year}{2021}\natexlab{}.
\newblock \bibinfo{title}{Scientific Research Methods}.
\newblock \bibinfo{howpublished}{\url{https://bookdown.org/pkaldunn/Book/}}.
\newblock


\bibitem[\protect\citeauthoryear{Durumeric, Bailey, and Halderman}{Durumeric
  et~al\mbox{.}}{2014}]%
        {durumeric2014internet}
\bibfield{author}{\bibinfo{person}{Zakir Durumeric}, \bibinfo{person}{Michael
  Bailey}, {and} \bibinfo{person}{J~Alex Halderman}.}
  \bibinfo{year}{2014}\natexlab{}.
\newblock \showarticletitle{An Internet-wide view of Internet-wide scanning}.
  In \bibinfo{booktitle}{\emph{23rd $\{$USENIX$\}$ Security Symposium
  ($\{$USENIX$\}$ Security 14)}}. \bibinfo{publisher}{{USENIX}},
  \bibinfo{address}{San Diego, California, USA}, \bibinfo{pages}{65--78}.
\newblock
\urldef\tempurl%
\url{https://jhalderm.com/pub/papers/scanning-sec14.pdf}
\showURL{%
\tempurl}


\bibitem[\protect\citeauthoryear{{Economist Editors}}{{Economist
  Editors}}{2018}]%
        {Economist18a}
\bibfield{author}{\bibinfo{person}{{Economist Editors}}.}
  \bibinfo{year}{2018}\natexlab{}.
\newblock \showarticletitle{Why some countries are turning off the internet on
  exam days}.
\newblock \bibinfo{journal}{\emph{The Economist}} (\bibinfo{date}{July 5}
  \bibinfo{year}{2018}).
\newblock
\urldef\tempurl%
\url{https://www.economist.com/middle-east-and-africa/2018/07/05/why-some-countries-are-turning-off-the-internet-on-exam-days}
\showURL{%
\tempurl}
\newblock
\shownote{(Appeared in the Middle East and Africa print edition).}


\bibitem[\protect\citeauthoryear{Electric}{Electric}{2021}]%
        {he_looking_glass}
\bibfield{author}{\bibinfo{person}{Hurricane Electric}.}
  \bibinfo{year}{2021}\natexlab{}.
\newblock \bibinfo{title}{Looking Glass}.
\newblock \bibinfo{howpublished}{\url{http://lg.he.net/}}.
\newblock


\bibitem[\protect\citeauthoryear{Engadget}{Engadget}{2020}]%
        {huawei2020}
\bibfield{author}{\bibinfo{person}{Engadget}.} \bibinfo{year}{2020}\natexlab{}.
\newblock \bibinfo{title}{{China}, {Huawei} propose internet protocol with a
  built-in killswitch}.
\newblock
  \bibinfo{howpublished}{\url{https://www.engadget.com/2020-03-30-china-huawei-new-ip-proposal.html}}.
\newblock


\bibitem[\protect\citeauthoryear{{Federal Networking Council (FNC)}}{{Federal
  Networking Council (FNC)}}{1995}]%
        {nitrd}
\bibfield{author}{\bibinfo{person}{{Federal Networking Council (FNC)}}.}
  \bibinfo{year}{1995}\natexlab{}.
\newblock \bibinfo{title}{{Definition of ``Internet''}}.
\newblock
  \bibinfo{howpublished}{\url{https://www.nitrd.gov/historical/fnc/internet_res.pdf}}.
\newblock


\bibitem[\protect\citeauthoryear{forums}{forums}{2017}]%
        {cloudflare_he}
\bibfield{author}{\bibinfo{person}{{HE} forums}.}
  \bibinfo{year}{2017}\natexlab{}.
\newblock \bibinfo{title}{Cloudflare Blocked on Free Tunnels now?}
\newblock
  \bibinfo{howpublished}{\url{https://forums.he.net/index.php?topic=3805.0}}.
\newblock


\bibitem[\protect\citeauthoryear{Francis and Gummadi}{Francis and
  Gummadi}{2001}]%
        {Francis01b}
\bibfield{author}{\bibinfo{person}{Paul Francis} {and}
  \bibinfo{person}{Ramakrishnan Gummadi}.} \bibinfo{year}{2001}\natexlab{}.
\newblock \showarticletitle{{IPNL}: A {NAT}-extended internet architecture}. In
  \bibinfo{booktitle}{\emph{Proceedings of the {ACM} SIGCOMM Conference}}.
  \bibinfo{publisher}{{ACM}}, \bibinfo{address}{San Diego, CA, USA},
  \bibinfo{pages}{69--80}.
\newblock
\urldef\tempurl%
\url{https://doi.org/10.1145/964723.383065}
\showDOI{\tempurl}


\bibitem[\protect\citeauthoryear{Fuller, Lear, and Meyer}{Fuller
  et~al\mbox{.}}{2008}]%
        {Fuller08a}
\bibfield{author}{\bibinfo{person}{V. Fuller}, \bibinfo{person}{E. Lear}, {and}
  \bibinfo{person}{D. Meyer}.} \bibinfo{year}{2008}\natexlab{}.
\newblock \bibinfo{title}{Reclassifying 240/4 as usable unicast address space}.
   (\bibinfo{date}{March} \bibinfo{year}{2008}).
\newblock
\urldef\tempurl%
\url{https://datatracker.ietf.org/doc/html/draft-fuller-240space-02}
\showURL{%
\tempurl}
\newblock
\shownote{Work in progress (Internet draft draft-fuller-240space-02.txt).}


\bibitem[\protect\citeauthoryear{Gao}{Gao}{2001}]%
        {Gao01b}
\bibfield{author}{\bibinfo{person}{Lixin Gao}.}
  \bibinfo{year}{2001}\natexlab{}.
\newblock \showarticletitle{On Inferring Autonomous System Relationships in the
  {Internet}}.
\newblock \bibinfo{journal}{\emph{{ACM/IEEE} Transactions on Networking}}
  \bibinfo{volume}{9}, \bibinfo{number}{6} (\bibinfo{date}{Dec.}
  \bibinfo{year}{2001}), \bibinfo{pages}{733--745}.
\newblock
\urldef\tempurl%
\url{https://doi.org/10.1109/90.974527}
\showDOI{\tempurl}


\bibitem[\protect\citeauthoryear{Gibbs}{Gibbs}{1996}]%
        {Gibbs16a}
\bibfield{author}{\bibinfo{person}{Samuel Gibbs}.}
  \bibinfo{year}{1996}\natexlab{}.
\newblock \showarticletitle{{Iraq} shuts down the {Internet} to stop pupils
  cheating in exams}.
\newblock \bibinfo{journal}{\emph{The Guardian}} (\bibinfo{date}{18 May}
  \bibinfo{year}{1996}).
\newblock
\urldef\tempurl%
\url{https://www.theguardian.com/technology/2016/may/18/iraq-shuts-down-internet-to-stop-pupils-cheating-in-exams}
\showURL{%
\tempurl}


\bibitem[\protect\citeauthoryear{{GovTrack.us}}{{GovTrack.us}}{2020}]%
        {GovTrack20a}
\bibfield{author}{\bibinfo{person}{{GovTrack.us}}.}
  \bibinfo{year}{2020}\natexlab{}.
\newblock \bibinfo{title}{Unplug the Internet Kill Switch Act would eliminate a
  1942 law that could let the president shut down the internet}.
\newblock
  \bibinfo{howpublished}{\url{https://govtrackinsider.com/unplug-the-internet-kill-switch-act-would-eliminate-a-1942-law-that-could-let-the-president-shut-78326f0ef66c}}.
\newblock
\urldef\tempurl%
\url{https://govtrackinsider.com/unplug-the-internet-kill-switch-act-would-eliminate-a-1942-law-that-could-let-the-president-shut-78326f0ef66c}
\showURL{%
\tempurl}


\bibitem[\protect\citeauthoryear{Greenberg, Hamilton, Jain, Kandula, Kim,
  Lahiri, Maltz, and Pat}{Greenberg et~al\mbox{.}}{2009}]%
        {Greenberg09a}
\bibfield{author}{\bibinfo{person}{Albert Greenberg}, \bibinfo{person}{James~R.
  Hamilton}, \bibinfo{person}{Navendu Jain}, \bibinfo{person}{Srikanth
  Kandula}, \bibinfo{person}{Changhoon Kim}, \bibinfo{person}{Parantap Lahiri},
  \bibinfo{person}{David~A. Maltz}, {and} \bibinfo{person}{Parveen Pat}.}
  \bibinfo{year}{2009}\natexlab{}.
\newblock \showarticletitle{{VL2}: A Scalable and Flexible Data Center
  Network}. In \bibinfo{booktitle}{\emph{Proceedings of the {ACM} SIGCOMM
  Conference}}. \bibinfo{publisher}{{ACM}}, \bibinfo{address}{Barcelona,
  Spain}, \bibinfo{pages}{51--62}.
\newblock
\urldef\tempurl%
\url{http://ccr.sigcomm.org/online/files/p51.pdf}
\showURL{%
\tempurl}


\bibitem[\protect\citeauthoryear{Griffiths}{Griffiths}{2019}]%
        {Griffiths19a}
\bibfield{author}{\bibinfo{person}{James Griffiths}.}
  \bibinfo{year}{2019}\natexlab{}.
\newblock \showarticletitle{Democratic Republic of Congo internet shutdown
  shows how Chinese censorship tactics are spreading}.
\newblock \bibinfo{journal}{\emph{CNN}} (\bibinfo{date}{Jan.~2}
  \bibinfo{year}{2019}).
\newblock
\urldef\tempurl%
\url{https://edition.cnn.com/2019/01/02/africa/congo-internet-shutdown-china-intl/index.html}
\showURL{%
\tempurl}


\bibitem[\protect\citeauthoryear{Guillot, Fontugne, Winter, Merindol, King,
  Dainotti, and Pelsser}{Guillot et~al\mbox{.}}{2019}]%
        {guillot2019internet}
\bibfield{author}{\bibinfo{person}{Andreas Guillot}, \bibinfo{person}{Romain
  Fontugne}, \bibinfo{person}{Philipp Winter}, \bibinfo{person}{Pascal
  Merindol}, \bibinfo{person}{Alistair King}, \bibinfo{person}{Alberto
  Dainotti}, {and} \bibinfo{person}{Cristel Pelsser}.}
  \bibinfo{year}{2019}\natexlab{}.
\newblock \showarticletitle{Chocolatine: Outage Detection for Internet
  Background Radiation}. In \bibinfo{booktitle}{\emph{Proceedings of the IFIP
  International Workshop on Traffic Monitoring and Analysis}}.
  \bibinfo{publisher}{IFIP}, \bibinfo{address}{Paris, France}, 8.
\newblock
\urldef\tempurl%
\url{https://clarinet.u-strasbg.fr/~pelsser/publications/Guillot-chocolatine-TMA2019.pdf}
\showURL{%
\tempurl}


\bibitem[\protect\citeauthoryear{Guo and Heidemann}{Guo and Heidemann}{2018}]%
        {Guo18a}
\bibfield{author}{\bibinfo{person}{Hang Guo} {and} \bibinfo{person}{John
  Heidemann}.} \bibinfo{year}{2018}\natexlab{}.
\newblock \showarticletitle{Detecting {ICMP} Rate Limiting in the {Internet}}.
  In \bibinfo{booktitle}{\emph{Proceedings of the Passive and Active
  Measurement Workshop}}. \bibinfo{publisher}{Springer},
  \bibinfo{address}{Berlin, Germany}, \bibinfo{pages}{to appear}.
\newblock


\bibitem[\protect\citeauthoryear{Heidemann, Pradkin, Govindan, Papadopoulos,
  Bartlett, and Bannister}{Heidemann et~al\mbox{.}}{2008}]%
        {Heidemann08c}
\bibfield{author}{\bibinfo{person}{John Heidemann}, \bibinfo{person}{Yuri
  Pradkin}, \bibinfo{person}{Ramesh Govindan}, \bibinfo{person}{Christos
  Papadopoulos}, \bibinfo{person}{Genevieve Bartlett}, {and}
  \bibinfo{person}{Joseph Bannister}.} \bibinfo{year}{2008}\natexlab{}.
\newblock \showarticletitle{Census and Survey of the Visible {Internet}}. In
  \bibinfo{booktitle}{\emph{Proceedings of the ACM Internet Measurement
  Conference}}. \bibinfo{publisher}{{ACM}}, \bibinfo{address}{Vouliagmeni,
  Greece}, \bibinfo{pages}{169--182}.
\newblock
\urldef\tempurl%
\url{https://doi.org/10.1145/1452520.1452542}
\showDOI{\tempurl}


\bibitem[\protect\citeauthoryear{Henley}{Henley}{2018}]%
        {Henley18a}
\bibfield{author}{\bibinfo{person}{Jon Henley}.}
  \bibinfo{year}{2018}\natexlab{}.
\newblock \showarticletitle{Algeria blocks internet to prevent students
  cheating during exams}.
\newblock \bibinfo{journal}{\emph{The Guardian}} (\bibinfo{date}{22 June}
  \bibinfo{year}{2018}).
\newblock
\urldef\tempurl%
\url{https://www.theguardian.com/world/2018/jun/21/algeria-shuts-internet-prevent-cheating-school-exams}
\showURL{%
\tempurl}


\bibitem[\protect\citeauthoryear{{IANA}}{{IANA}}{2021}]%
        {iana_v6}
\bibfield{author}{\bibinfo{person}{{IANA}}.} \bibinfo{year}{2021}\natexlab{}.
\newblock \bibinfo{title}{IPv6 {RIR} Allocation Data}.
\newblock
  \bibinfo{howpublished}{\url{https://www.iana.org/numbers/allocations/}}.
\newblock


\bibitem[\protect\citeauthoryear{{ICANN}}{{ICANN}}{2011}]%
        {ICANN11a}
\bibfield{author}{\bibinfo{person}{{ICANN}}.} \bibinfo{year}{2011}\natexlab{}.
\newblock \bibinfo{booktitle}{\emph{Available Pool of Unallocated IPv4 Internet
  Addresses Now Completely Emptied}}.
\newblock \bibinfo{type}{Announcement}. \bibinfo{institution}{ICANN}.
\newblock
\urldef\tempurl%
\url{https://itp.cdn.icann.org/en/files/announcements/release-03feb11-en.pdf}
\showURL{%
\tempurl}


\bibitem[\protect\citeauthoryear{{Internet Architecture Board}}{{Internet
  Architecture Board}}{2000}]%
        {IAB20a}
\bibfield{author}{\bibinfo{person}{{Internet Architecture Board}}.}
  \bibinfo{year}{2000}\natexlab{}.
\newblock \bibinfo{booktitle}{\emph{{IAB} Technical Comment on the Unique {DNS}
  Root}}.
\newblock \bibinfo{type}{RFC} 2826. \bibinfo{institution}{Internet Request For
  Comments}.
\newblock
\urldef\tempurl%
\url{https://www.rfc-editor.org/rfc/rfc2826}
\showURL{%
\tempurl}


\bibitem[\protect\citeauthoryear{Katz-Bassett, Madhyastha, John, Krishnamurthy,
  Wetherall, and Anderson}{Katz-Bassett et~al\mbox{.}}{2008}]%
        {katz2008studying}
\bibfield{author}{\bibinfo{person}{Ethan Katz-Bassett},
  \bibinfo{person}{Harsha~V Madhyastha}, \bibinfo{person}{John~P John},
  \bibinfo{person}{Arvind Krishnamurthy}, \bibinfo{person}{David Wetherall},
  {and} \bibinfo{person}{Thomas~E Anderson}.} \bibinfo{year}{2008}\natexlab{}.
\newblock \showarticletitle{Studying Black Holes in the Internet with Hubble}.
  In \bibinfo{booktitle}{\emph{Proceedings of the USENIX Conference on
  Networked Systems Design and Implementation}}. \bibinfo{publisher}{ACM},
  \bibinfo{address}{San Francisco, CA}, \bibinfo{pages}{247--262}.
\newblock


\bibitem[\protect\citeauthoryear{Katz-Bassett, Scott, Choffnes, Cunha,
  Valancius, Feamster, Madhyastha, Anderson, and Krishnamurthy}{Katz-Bassett
  et~al\mbox{.}}{2012}]%
        {katz2012lifeguard}
\bibfield{author}{\bibinfo{person}{Ethan Katz-Bassett}, \bibinfo{person}{Colin
  Scott}, \bibinfo{person}{David~R. Choffnes}, \bibinfo{person}{{\'I}talo
  Cunha}, \bibinfo{person}{Vytautas Valancius}, \bibinfo{person}{Nick
  Feamster}, \bibinfo{person}{Harsha~V. Madhyastha}, \bibinfo{person}{Tom
  Anderson}, {and} \bibinfo{person}{Arvind Krishnamurthy}.}
  \bibinfo{year}{2012}\natexlab{}.
\newblock \showarticletitle{{LIFEGUARD}: Practical Repair of Persistent Route
  Failures}. In \bibinfo{booktitle}{\emph{Proceedings of the {ACM} SIGCOMM
  Conference}}. \bibinfo{publisher}{{ACM}}, \bibinfo{address}{Helsinki,
  Finland}, \bibinfo{pages}{395--406}.
\newblock
\urldef\tempurl%
\url{https://doi.org/10.1145/2377677.2377756}
\showDOI{\tempurl}


\bibitem[\protect\citeauthoryear{Knowledge}{Knowledge}{2009}]%
        {ipv6peeringdisputes}
\bibfield{author}{\bibinfo{person}{DataCenter Knowledge}.}
  \bibinfo{year}{2009}\natexlab{}.
\newblock \bibinfo{title}{{Peering Disputes Migrate to IPv6}}.
\newblock
  \bibinfo{howpublished}{\url{https://www.datacenterknowledge.com/archives/2009/10/22/peering-disputes-migrate-to-ipv6}}.
\newblock


\bibitem[\protect\citeauthoryear{Lamport}{Lamport}{1998}]%
        {Lamport98a}
\bibfield{author}{\bibinfo{person}{Leslie Lamport}.}
  \bibinfo{year}{1998}\natexlab{}.
\newblock \showarticletitle{The Part-Time Parliament}.
\newblock \bibinfo{journal}{\emph{{ACM} Transactions on Computer Systems}}
  \bibinfo{volume}{16}, \bibinfo{number}{2} (\bibinfo{date}{May}
  \bibinfo{year}{1998}), \bibinfo{pages}{133--169}.
\newblock
\urldef\tempurl%
\url{https://doi.org/10.1145/279227.279229}
\showDOI{\tempurl}


\bibitem[\protect\citeauthoryear{Lamport, Shostak, and Pease}{Lamport
  et~al\mbox{.}}{1982}]%
        {Lamport82a}
\bibfield{author}{\bibinfo{person}{Leslie Lamport}, \bibinfo{person}{Robert
  Shostak}, {and} \bibinfo{person}{Marshall Pease}.}
  \bibinfo{year}{1982}\natexlab{}.
\newblock \showarticletitle{The {Byzantine} Generals Problem}.
\newblock \bibinfo{journal}{\emph{{ACM} Transactions on Programming Languages
  and Systems}} \bibinfo{volume}{4}, \bibinfo{number}{3} (\bibinfo{date}{July}
  \bibinfo{year}{1982}), \bibinfo{pages}{382--401}.
\newblock


\bibitem[\protect\citeauthoryear{Lemley and Lessig}{Lemley and Lessig}{2001}]%
        {Lemley01a}
\bibfield{author}{\bibinfo{person}{Mark~A. Lemley} {and}
  \bibinfo{person}{Lawrence Lessig}.} \bibinfo{year}{2001}\natexlab{}.
\newblock \showarticletitle{The End of End-to-End: Preserving the Architecture
  of the Internet in the Broadband Era}.
\newblock \bibinfo{journal}{\emph{UCLA Law Review}}  \bibinfo{volume}{48}
  (\bibinfo{date}{April} \bibinfo{year}{2001}), \bibinfo{pages}{925--972}.
\newblock
\urldef\tempurl%
\url{https://law.stanford.edu/publications/the-end-of-end-to-end-preserving-the-architecture-of-the-internet-in-the-broadband-era/}
\showURL{%
\tempurl}


\bibitem[\protect\citeauthoryear{Luckie, Huffaker, Dhamdhere, Giotsas, and
  kc~claffy}{Luckie et~al\mbox{.}}{2013}]%
        {Luckie13a}
\bibfield{author}{\bibinfo{person}{Matthew Luckie}, \bibinfo{person}{Bradley
  Huffaker}, \bibinfo{person}{Dhamdhere}, \bibinfo{person}{Vasileios Giotsas},
  {and} \bibinfo{person}{kc claffy}.} \bibinfo{year}{2013}\natexlab{}.
\newblock \showarticletitle{{AS} Relationships, Customer Cones, and
  Validation}. In \bibinfo{booktitle}{\emph{Proceedings of the ACM Internet
  Measurement Conference}}. \bibinfo{publisher}{{ACM}},
  \bibinfo{address}{Barcelona, Spain}, \bibinfo{pages}{243--256}.
\newblock
\urldef\tempurl%
\url{http://conferences.sigcomm.org/imc/2013/papers/imc039-luckieAemb.pdf}
\showURL{%
\tempurl}


\bibitem[\protect\citeauthoryear{Miller, Nixon, Tai, and Wood}{Miller
  et~al\mbox{.}}{2001}]%
        {Miller01a}
\bibfield{author}{\bibinfo{person}{Brent~A. Miller}, \bibinfo{person}{Toby
  Nixon}, \bibinfo{person}{Charlie Tai}, {and} \bibinfo{person}{Mark~D. Wood}.}
  \bibinfo{year}{2001}\natexlab{}.
\newblock \showarticletitle{Home Networking with Universal Plug and Play}.
\newblock \bibinfo{journal}{\emph{{IEEE} Communications Magazine}}
  \bibinfo{volume}{39}, \bibinfo{number}{12} (\bibinfo{date}{Dec.}
  \bibinfo{year}{2001}), \bibinfo{pages}{104--109}.
\newblock
\urldef\tempurl%
\url{https://doi.org/10.1109/35.968819}
\showDOI{\tempurl}


\bibitem[\protect\citeauthoryear{Miller}{Miller}{2009}]%
        {Miller09a}
\bibfield{author}{\bibinfo{person}{Rich Miller}.}
  \bibinfo{year}{2009}\natexlab{}.
\newblock \bibinfo{title}{Peering Disputes Migrate to {IPv6}}.
\newblock \bibinfo{howpublished}{website
  \url{https://www.datacenterknowledge.com/archives/2009/10/22/peering-disputes-migrate-to-ipv6}}.
\newblock
\urldef\tempurl%
\url{https://www.datacenterknowledge.com/archives/2009/10/22/peering-disputes-migrate-to-ipv6}
\showURL{%
\tempurl}


\bibitem[\protect\citeauthoryear{Nakamoto}{Nakamoto}{2009}]%
        {Nakamoto09a}
\bibfield{author}{\bibinfo{person}{Satoshi Nakamoto}.}
  \bibinfo{year}{2009}\natexlab{}.
\newblock \bibinfo{title}{Bitcoin: A Peer-to-Peer Electronic Cash System}.
\newblock \bibinfo{howpublished}{Released publically
  \url{http://bitcoin.org/bitcoin.pdf}}.
\newblock
\urldef\tempurl%
\url{http://bitcoin.org/bitcoin.pdf}
\showURL{%
\tempurl}


\bibitem[\protect\citeauthoryear{News}{News}{2019}]%
        {BBC19a}
\bibfield{author}{\bibinfo{person}{{BBC} News}.}
  \bibinfo{year}{2019}\natexlab{}.
\newblock \bibinfo{title}{Russia internet: Law introducing new controls comes
  into force}.
\newblock \bibinfo{howpublished}{website
  \url{https://www.bbc.com/news/world-europe-50259597}}.
\newblock
\urldef\tempurl%
\url{https://www.bbc.com/news/world-europe-50259597}
\showURL{%
\tempurl}


\bibitem[\protect\citeauthoryear{{NRO}}{{NRO}}{2021}]%
        {iana_v4}
\bibfield{author}{\bibinfo{person}{{NRO}}.} \bibinfo{year}{2021}\natexlab{}.
\newblock \bibinfo{title}{IPv4 Address Space Registry}.
\newblock
  \bibinfo{howpublished}{\url{https://www.nro.net/about/rirs/statistics/}}.
\newblock


\bibitem[\protect\citeauthoryear{Partridge, Mendez, and Milliken}{Partridge
  et~al\mbox{.}}{1993}]%
        {Partridge93a}
\bibfield{author}{\bibinfo{person}{C. Partridge}, \bibinfo{person}{T. Mendez},
  {and} \bibinfo{person}{W. Milliken}.} \bibinfo{year}{1993}\natexlab{}.
\newblock \bibinfo{booktitle}{\emph{Host Anycasting Service}}.
\newblock \bibinfo{type}{RFC} 1546. \bibinfo{institution}{Internet Request For
  Comments}.
\newblock
\urldef\tempurl%
\url{https://www.rfc-editor.org/rfc/rfc1546.txt}
\showURL{%
\tempurl}


\bibitem[\protect\citeauthoryear{Postel}{Postel}{1980}]%
        {Postel80b}
\bibfield{author}{\bibinfo{person}{Jonathan~B. Postel}.}
  \bibinfo{year}{1980}\natexlab{}.
\newblock \showarticletitle{{Internetwork} Protocol Approaches}.
\newblock \bibinfo{journal}{\emph{IEEE Trans. Comput.}} \bibinfo{volume}{28},
  \bibinfo{number}{4} (\bibinfo{date}{April} \bibinfo{year}{1980}),
  \bibinfo{pages}{604--611}.
\newblock
\urldef\tempurl%
\url{https://doi.org/10.1109/TCOM.1980.1094705}
\showDOI{\tempurl}


\bibitem[\protect\citeauthoryear{Quan, Heidemann, and Pradkin}{Quan
  et~al\mbox{.}}{2013}]%
        {quan2013trinocular}
\bibfield{author}{\bibinfo{person}{Lin Quan}, \bibinfo{person}{John Heidemann},
  {and} \bibinfo{person}{Yuri Pradkin}.} \bibinfo{year}{2013}\natexlab{}.
\newblock \showarticletitle{Trinocular: Understanding {Internet} Reliability
  Through Adaptive Probing}. In \bibinfo{booktitle}{\emph{Proceedings of the
  {ACM} SIGCOMM Conference}}. \bibinfo{publisher}{{ACM}},
  \bibinfo{address}{Hong Kong, China}, \bibinfo{pages}{255--266}.
\newblock
\urldef\tempurl%
\url{https://doi.org/10.1145/2486001.2486017}
\showDOI{\tempurl}


\bibitem[\protect\citeauthoryear{Rayburn}{Rayburn}{2016}]%
        {google_cogent}
\bibfield{author}{\bibinfo{person}{Dan Rayburn}.}
  \bibinfo{year}{2016}\natexlab{}.
\newblock \bibinfo{title}{Google Blocking IPv6 Adoption With Cogent, Impacting
  Transit Customers}.
\newblock
  \bibinfo{howpublished}{\url{https://seekingalpha.com/article/3948876-google-blocking-ipv6-adoption-cogent-impacting-transit-customers}}.
\newblock


\bibitem[\protect\citeauthoryear{Reuters}{Reuters}{2021}]%
        {Reuters21a}
\bibfield{author}{\bibinfo{person}{Reuters}.} \bibinfo{year}{2021}\natexlab{}.
\newblock \bibinfo{title}{Russia disconnected from internet in tests as it
  bolsters security}.
\newblock \bibinfo{howpublished}{website
  \url{https://www.reuters.com/technology/russia-disconnected-global-internet-tests-rbc-daily-2021-07-22/}}.
\newblock
\urldef\tempurl%
\url{https://www.reuters.com/technology/russia-disconnected-global-internet-tests-rbc-daily-2021-07-22/}
\showURL{%
\tempurl}


\bibitem[\protect\citeauthoryear{Reuters}{Reuters}{2022}]%
        {Reuters22a}
\bibfield{author}{\bibinfo{person}{Reuters}.} \bibinfo{year}{2022}\natexlab{}.
\newblock \bibinfo{howpublished}{website
  \url{https://www.reuters.com/technology/us-firm-cogent-cutting-internet-service-russia-2022-03-04/}}.
\newblock
\urldef\tempurl%
\url{https://www.reuters.com/technology/us-firm-cogent-cutting-internet-service-russia-2022-03-04/}
\showURL{%
\tempurl}


\bibitem[\protect\citeauthoryear{Richter, Padmanabhan, Spring, Berger, and
  Clark}{Richter et~al\mbox{.}}{2018}]%
        {richter2018advancing}
\bibfield{author}{\bibinfo{person}{Philipp Richter},
  \bibinfo{person}{Ramakrishna Padmanabhan}, \bibinfo{person}{Neil Spring},
  \bibinfo{person}{Arthur Berger}, {and} \bibinfo{person}{David Clark}.}
  \bibinfo{year}{2018}\natexlab{}.
\newblock \showarticletitle{Advancing the Art of {Internet} Edge Outage
  Detection}. In \bibinfo{booktitle}{\emph{Proceedings of the ACM Internet
  Measurement Conference}}. \bibinfo{publisher}{{ACM}},
  \bibinfo{address}{Boston, Massachusetts, USA}, \bibinfo{pages}{350--363}.
\newblock
\urldef\tempurl%
\url{https://doi.org/10.1145/3278532.3278563}
\showDOI{\tempurl}


\bibitem[\protect\citeauthoryear{Richter, Wohlfart, Vallina-Rodriguez, Allman,
  Bush, Feldmann, Kreibich, Weaver, and Paxson}{Richter et~al\mbox{.}}{2016}]%
        {Richter16c}
\bibfield{author}{\bibinfo{person}{Philipp Richter}, \bibinfo{person}{Florian
  Wohlfart}, \bibinfo{person}{Narseo Vallina-Rodriguez}, \bibinfo{person}{Mark
  Allman}, \bibinfo{person}{Randy Bush}, \bibinfo{person}{Anja Feldmann},
  \bibinfo{person}{Christian Kreibich}, \bibinfo{person}{Nicholas Weaver},
  {and} \bibinfo{person}{Vern Paxson}.} \bibinfo{year}{2016}\natexlab{}.
\newblock \showarticletitle{A Multi-perspective Analysis of Carrier-Grade {NAT}
  Deployment}. In \bibinfo{booktitle}{\emph{Proceedings of the ACM Internet
  Measurement Conference}}. \bibinfo{publisher}{{ACM}}, \bibinfo{address}{Santa
  Monica, CA, USA}.
\newblock
\urldef\tempurl%
\url{https://doi.org/10.1145/2987443.2987474}
\showDOI{\tempurl}


\bibitem[\protect\citeauthoryear{{RIPE NCC Staff}}{{RIPE NCC Staff}}{2015}]%
        {Ripe15c}
\bibfield{author}{\bibinfo{person}{{RIPE NCC Staff}}.}
  \bibinfo{year}{2015}\natexlab{}.
\newblock \showarticletitle{{RIPE} {Atlas}: A Global {Internet} Measurement
  Network}.
\newblock \bibinfo{journal}{\emph{The Internet Protocol Journal}}
  \bibinfo{volume}{18}, \bibinfo{number}{3} (\bibinfo{date}{Sept.}
  \bibinfo{year}{2015}), \bibinfo{pages}{2--26}.
\newblock


\bibitem[\protect\citeauthoryear{Rockefeller}{Rockefeller}{2009}]%
        {cybersecurity_act_2010}
\bibfield{author}{\bibinfo{person}{Sen.~John~D. Rockefeller}.}
  \bibinfo{year}{2009}\natexlab{}.
\newblock \bibinfo{title}{Cybersecurity Act of 2010}.
\newblock
  \bibinfo{howpublished}{\url{https://www.congress.gov/bill/111th-congress/senate-bill/773}}.
\newblock


\bibitem[\protect\citeauthoryear{{Root Operators}}{{Root Operators}}{2016}]%
        {RootServers16a}
\bibfield{author}{\bibinfo{person}{{Root Operators}}.}
  \bibinfo{year}{2016}\natexlab{}.
\newblock \bibinfo{title}{\url{http://www.root-servers.org}}.
\newblock
\newblock


\bibitem[\protect\citeauthoryear{Rosenberg, Weinberger, Huitema, and
  Mahy}{Rosenberg et~al\mbox{.}}{2003}]%
        {Rosenberg03a}
\bibfield{author}{\bibinfo{person}{J. Rosenberg}, \bibinfo{person}{J.
  Weinberger}, \bibinfo{person}{C. Huitema}, {and} \bibinfo{person}{R. Mahy}.}
  \bibinfo{year}{2003}\natexlab{}.
\newblock \bibinfo{booktitle}{\emph{{STUN}---Simple Traversal of User Datagram
  Protocol ({UDP}) Through Network Address Translators ({NATs})}}.
\newblock \bibinfo{type}{RFC} 3489. \bibinfo{institution}{Internet Request For
  Comments}.
\newblock
\urldef\tempurl%
\url{ftp://ftp.rfc-editor.org/in-notes/rfc3489.txt}
\showURL{%
\tempurl}


\bibitem[\protect\citeauthoryear{Saluja, Heidemann, and Pradkin}{Saluja
  et~al\mbox{.}}{2022}]%
        {Saluja22a}
\bibfield{author}{\bibinfo{person}{Tarang Saluja}, \bibinfo{person}{John
  Heidemann}, {and} \bibinfo{person}{Yuri Pradkin}.}
  \bibinfo{year}{2022}\natexlab{}.
\newblock \showarticletitle{Differences in Monitoring the {DNS} Root Over
  {IPv4} and {IPv6}}. In \bibinfo{booktitle}{\emph{Proceedings of the National
  Symposium for NSF REU Research in Data Science, Systems, and Security}}.
  \bibinfo{publisher}{{IEEE}}, \bibinfo{address}{Portland, OR, USA},
  \bibinfo{pages}{to appear}.
\newblock


\bibitem[\protect\citeauthoryear{Schlinker, Kim, Cui, Katz-Bassett, Madhyastha,
  Cunha, Quinn, Hasan, Lapukhov, and Zeng}{Schlinker et~al\mbox{.}}{2017}]%
        {Schlinker17a}
\bibfield{author}{\bibinfo{person}{Brandon Schlinker},
  \bibinfo{person}{Hyojeong Kim}, \bibinfo{person}{Timothy Cui},
  \bibinfo{person}{Ethan Katz-Bassett}, \bibinfo{person}{Harsha~V. Madhyastha},
  \bibinfo{person}{Italo Cunha}, \bibinfo{person}{James Quinn},
  \bibinfo{person}{Saif Hasan}, \bibinfo{person}{Petr Lapukhov}, {and}
  \bibinfo{person}{Hongyi Zeng}.} \bibinfo{year}{2017}\natexlab{}.
\newblock \showarticletitle{Engineering Egress with {Edge} {Fabric}: Steering
  Oceans of Content to the World}. In \bibinfo{booktitle}{\emph{Proceedings of
  the {ACM} SIGCOMM Conference}}. \bibinfo{publisher}{{ACM}},
  \bibinfo{address}{Los Angeles, CA, USA}, \bibinfo{pages}{418--431}.
\newblock
\urldef\tempurl%
\url{https://doi.org/10.1145/3098822.3098853}
\showDOI{\tempurl}


\bibitem[\protect\citeauthoryear{Schulman and Spring}{Schulman and
  Spring}{2011}]%
        {Schulman11a}
\bibfield{author}{\bibinfo{person}{Aaron Schulman} {and} \bibinfo{person}{Neil
  Spring}.} \bibinfo{year}{2011}\natexlab{}.
\newblock \showarticletitle{Pingin' in the Rain}. In
  \bibinfo{booktitle}{\emph{Proceedings of the ACM Internet Measurement
  Conference}}. \bibinfo{publisher}{{ACM}}, \bibinfo{address}{Berlin, Germany},
  \bibinfo{pages}{19--25}.
\newblock
\urldef\tempurl%
\url{https://doi.org/10.1145/2068816.2068819}
\showDOI{\tempurl}


\bibitem[\protect\citeauthoryear{Shah, Fontugne, Aben, Pelsser, and Bush}{Shah
  et~al\mbox{.}}{2017}]%
        {Shah17a}
\bibfield{author}{\bibinfo{person}{Anant Shah}, \bibinfo{person}{Romain
  Fontugne}, \bibinfo{person}{Emile Aben}, \bibinfo{person}{Cristel Pelsser},
  {and} \bibinfo{person}{Randy Bush}.} \bibinfo{year}{2017}\natexlab{}.
\newblock \showarticletitle{Disco: Fast, Good, and Cheap Outage Detection}. In
  \bibinfo{booktitle}{\emph{Proceedings of the IEEE International Conference on
  Traffic Monitoring and Analysis}}. \bibinfo{publisher}{Springer},
  \bibinfo{address}{Dublin, Ireland}, \bibinfo{pages}{1--9}.
\newblock
\urldef\tempurl%
\url{https://doi.org/10.23919/TMA.2017.8002902}
\showDOI{\tempurl}


\bibitem[\protect\citeauthoryear{Taye and Cheng}{Taye and Cheng}{2019}]%
        {Taye19a}
\bibfield{author}{\bibinfo{person}{Berhan Taye} {and} \bibinfo{person}{Sage
  Cheng}.} \bibinfo{year}{2019}\natexlab{}.
\newblock \bibinfo{title}{Report: the state of internet shutdowns}.
\newblock \bibinfo{howpublished}{blog
  \url{https://www.accessnow.org/the-state-of-internet-shutdowns-in-2018/}}.
\newblock
\urldef\tempurl%
\url{https://www.accessnow.org/the-state-of-internet-shutdowns-in-2018/}
\showURL{%
\tempurl}


\bibitem[\protect\citeauthoryear{Timberg and Sonne}{Timberg and Sonne}{2021}]%
        {Timberg21a}
\bibfield{author}{\bibinfo{person}{Craig Timberg} {and} \bibinfo{person}{Paul
  Sonne}.} \bibinfo{year}{2021}\natexlab{}.
\newblock \showarticletitle{Minutes before {Trump} left office, millions of the
  {Pentagon's} dormant {IP} addresses sprang to life}.
\newblock \bibinfo{journal}{\emph{The Washington Post}}
  (\bibinfo{date}{Apr.~24} \bibinfo{year}{2021}).
\newblock
\urldef\tempurl%
\url{https://www.washingtonpost.com/technology/2021/04/24/pentagon-internet-address-mystery/}
\showURL{%
\tempurl}


\bibitem[\protect\citeauthoryear{Tsuchiya and Eng}{Tsuchiya and Eng}{1993}]%
        {Tsuchiya93a}
\bibfield{author}{\bibinfo{person}{Paul~F. Tsuchiya} {and}
  \bibinfo{person}{Tony Eng}.} \bibinfo{year}{1993}\natexlab{}.
\newblock \showarticletitle{Extending the {IP} {Internet} Through Address
  Reuse}.
\newblock \bibinfo{journal}{\emph{{ACM} Computer Communication Review}}
  \bibinfo{volume}{23}, \bibinfo{number}{1} (\bibinfo{date}{Jan.}
  \bibinfo{year}{1993}), \bibinfo{pages}{16--33}.
\newblock
\urldef\tempurl%
\url{http://www.cs.cornell.edu/People/francis/tsuchiya93extending.pdf}
\showURL{%
\tempurl}


\bibitem[\protect\citeauthoryear{{USC/LANDER Project}}{{USC/LANDER
  Project}}{2017}]%
        {LANDER17a}
\bibfield{author}{\bibinfo{person}{{USC/LANDER Project}}.}
  \bibinfo{year}{2017}\natexlab{}.
\newblock \bibinfo{title}{Internet Outage Measurements}.
\newblock \bibinfo{howpublished}{IMPACT ID
  \url{USC-LANDER/LANDER:internet_outage_adaptive_a30all-20171006} at
  \url{https://ant.isi.edu/datasets/internet_outages/}}.
\newblock


\bibitem[\protect\citeauthoryear{Zander, Andrew, and Armitage}{Zander
  et~al\mbox{.}}{2014}]%
        {Zander14b}
\bibfield{author}{\bibinfo{person}{Sebastian Zander}, \bibinfo{person}{Lachlan
  L.~H. Andrew}, {and} \bibinfo{person}{Grenville Armitage}.}
  \bibinfo{year}{2014}\natexlab{}.
\newblock \showarticletitle{Capturing Ghosts: Predicting the Used {IPv4} Space
  by Inferring Unobserved Addresses}. In \bibinfo{booktitle}{\emph{Proceedings
  of the ACM Internet Measurement Conference}}. \bibinfo{publisher}{{ACM}},
  \bibinfo{address}{Vancouver, BC, Canada}, \bibinfo{pages}{319--332}.
\newblock
\urldef\tempurl%
\url{https://doi.org/10.1145/2663716.2663718}
\showDOI{\tempurl}


\end{thebibliography}

\appendix
\section{Examples}
	\label{sec:peninsula_exampla_data}

In \autoref{sec:internet_landscape}
  we define islands and peninsulas as two cases of partial connectivity
  in the Internet.
Here we give two real-world examples of islands and peninsulas that we
  discovered in Trinocular data.

For our example island (\autoref{fig:a28all_417bca00_accum})
  and peninusla (\autoref{fig:a30all_50f5b000_accum}),
  we show a graph that counts
  the number of active IP addresses that are reachable
  from the 6 Trinocular observers.
Each line represents the best estimate of
  the current number of active addresses from each observer.
Most of the lines generally overlap and show a V-shaped
  dip during the island,
  but one (e, the green line),
  stands out as fairly stable over this period.
Because we probe only a few addresses per round,
  the estimate of active addresses updates slowly,
  and lags the true value after an abrupt change in reachability.

While \emph{visually} these graphs look the same, with one VP able to reach
  the destination block while all others fail,
  we distinguish the island from the peninsulas with additional information.
For the example island, the destination block also hosts VP E,
  so we know with confidence E can reach itself
  but cannot reach the rest of the Internet, making it an island.
For the example peninsula,
  all VPs are external to the unreachable block.
We confirm that it is a peninsula using BGP and traceroutes.
We can therefore distinguish them as island and peninsula
  both from the Trinocular observations,
  and to confirm that claim with external data sources.

\subsection{An Example Island}

We defined island in \autoref{sec:island_definition},
  and looked for the systematically in Trinocular data
  (as described
  elsewhere~\cite{Baltra23a}).

\begin{figure}
\includegraphics[width=\columnwidth]{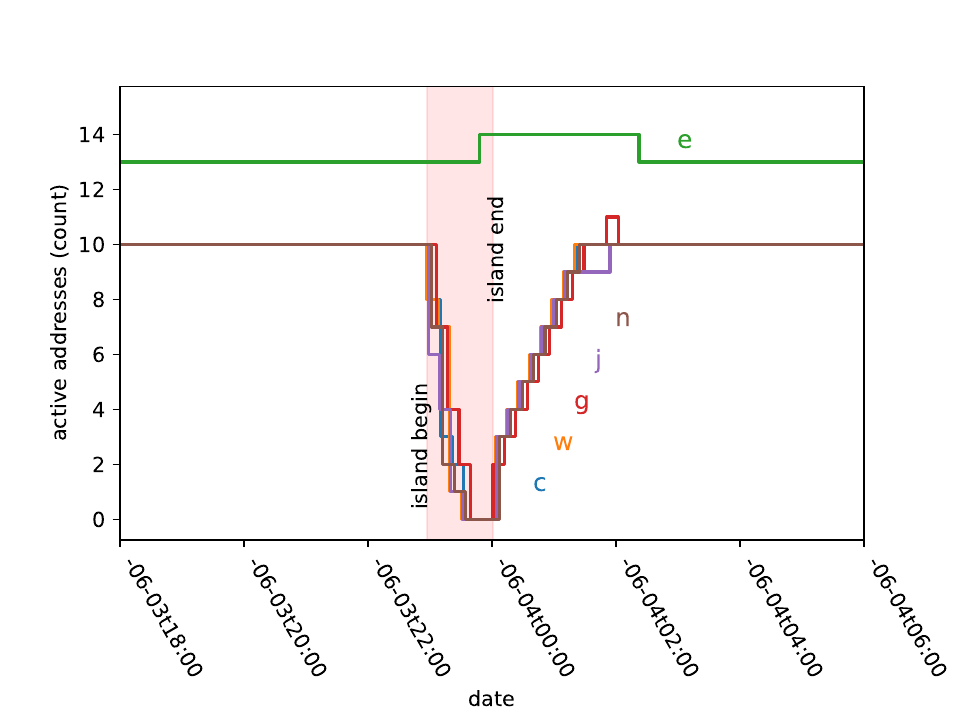}
\caption{Estimates of an island in Trinocular data
  starting 2017-06-03t23:06Z and lasting just longer than one hour.}
    \label{fig:a28all_417bca00_accum}
\end{figure}

\autoref{fig:a28all_417bca00_accum} shows an example island we discovered.
This island begins at 2017-06-03t23:06Z and lasts about 64 minutes.

For this hour, VP E and this network are part of an island, cut off from
  the rest of the internet and other VPs.
Because VP ``E'' is inside the island,
  it always sees 13 (or 14) active IP addresses
  in its top green line \autoref{fig:a28all_417bca00_accum}.
By contrast, the other 5 sites see a steady-state of 10 VPs,
  dropping to 0 during the island,
  as the gradually re-scan and fail to reach previously active IP addresses.
Other VPs rediscover all 10 addresses
  over the next two hours after the island ends.
(VP E sees 3 more IP active addresses that the other VPs for this block
  presumably because those targets have firewall rules that only
  allow replies to sources originating from the same block.)

The network being scanned here was the same network block hosting VP E,
  and we confirmed that this network was disconnected from network operators.
During the hour-long island VP E had 5 Trinocular rounds to scan the whole Internet,
  and it concluded that about 80\% of the Internet was unreachable.
It actually had failures to all of the Internet,
  but it dismisses brief unreachability to 20\% of blocks
  due to a conservative choice from the FBS algorithm~\cite{Baltra20a}.
(This algorithm requires that outages in sparsely active blocks
  are only confirmed after all probed addresses in the block
  respond negatively.
Incremental scanning all addresses in about 20\% of
  blocks takes longer than one hour in this dataset.)

\subsection{An Example Peninusla}

We defined peninsulas in \autoref{sec:peninsula_definition},
  and discussed an example that occurred in 2017-10-23 in Poland.

\begin{figure}
\includegraphics[width=\columnwidth]{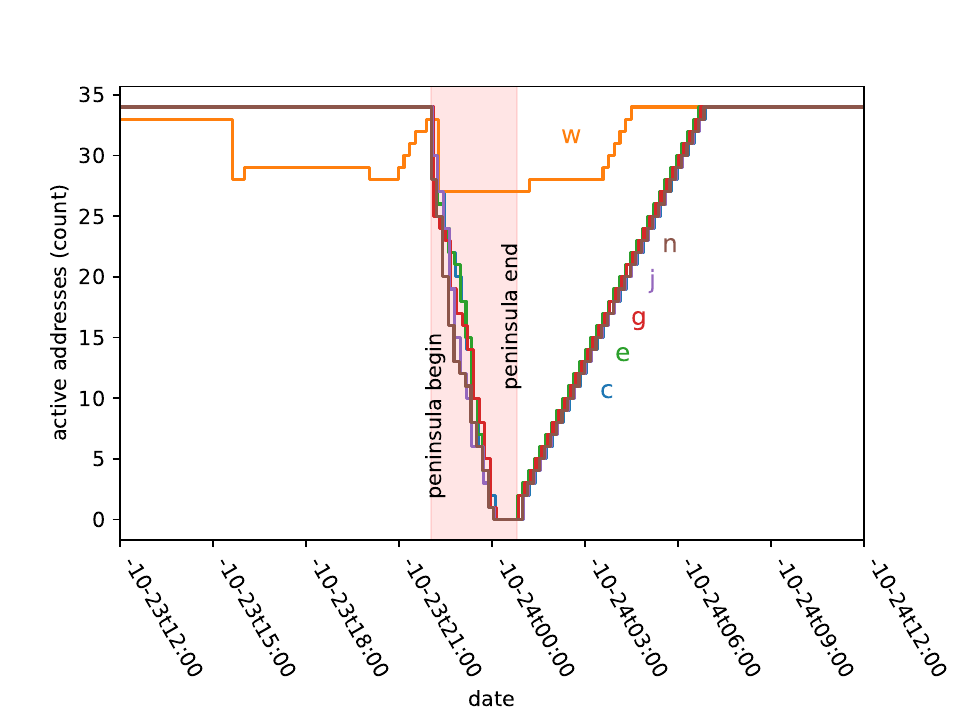}
\caption{Estimates of reachable addresses during a peninsula found in Trinocular data
  starting 2017-10-23t22:02Z and lasting about 3 hours.}
    \label{fig:a30all_50f5b000_accum}
\end{figure}

This peninsula was discovered by algorithms we developed
  (described
  elsewhere~\cite{Baltra23a}).
\autoref{fig:a30all_50f5b000_accum} shows
  our best estimates of the number of responsive addresses
  from each of our 6 observers.
We believe this block has 78 active addresses (based on about 3 years of
  history), and we scan a mean of 2.3 addresses each 11-minute round.

We believe the peninsula begins
  at 2017-10-23t22:02:24,
  the time the first observer (n) has no successful queries,
  and ends 2h46m later
  when another observer (e) is successful.
The w observer is successful for the whole period,
  as shown by its orange line staying fixed at 27
  during the peninsula,
  confirming that its queries sent reach a responsive address every 660\,s.
By contrast, the other five observers
  have no positive responses during the peninsula,
  so their active address count drops to zero
  by midnight.
After the peninsula ends,
  queries the five sites are again successful
  and the estimate of active addresses climbs slow back to 34
  just after t06:00.

To explain why the observed number of active addresses
  lags true reachability,
  recall that, in each round,
  Trinocular probes the minimum number of addresses
  to confirm block reachability.
After the peninsula recovers,
  this means one address each round confirms reachability,
  explaining the the 33 rounds (5.5\,h) it takes to return to
  34 active addresses.
(Each site actually probes two addresses in
  the round just after recovery.)
Negative information is acquired more quickly,
  because each site requires 3 to 5 negative responses
  to confirm unreachability during the peninsula.
Although Trinocular is willing to send up to 15 queries per round,
  this block usually full and stable
  (typically 34 of 78 addresses responding, giving an expected
  response around around $A=0.44$),
  so it can conclude unreachability from fewer queries.

This example is representative of other peninsulas we have seen.

\label{page:last_page}

\end{document}